\documentclass{ieeeaccess}
\usepackage{cite}
\usepackage{amsmath,amssymb,amsfonts}
\usepackage{algorithmic}
\usepackage{graphicx}
\usepackage{textcomp}
\usepackage{wrapfig}

\usepackage{color}
\usepackage{mathtools}
\usepackage{caption}
\usepackage{multirow}
\usepackage{bm,bbm}
\usepackage{booktabs}
\usepackage[caption=false,font=footnotesize]{subfig}
\usepackage[whole]{bxcjkjatype}
\graphicspath{{./img/}}
\usepackage{textcomp}
\usepackage{mathcomp}

\def\BibTeX{{\rm B\kern-.05em{\sc i\kern-.025em b}\kern-.08em
    T\kern-.1667em\lower.7ex\hbox{E}\kern-.125emX}}

\begin{document}

\history{Date of publication xxxx 00, 0000, date of current version xxxx 00, 0000.}
\doi{10.1109/ACCESS.2017.DOI}

\title{Bi-directional Beamforming Feedback-based Firmware-agnostic WiFi Sensing: An Empirical Study}
\author{
    \uppercase{Sota~Kondo}\authorrefmark{1}, \IEEEmembership{Student~Member,~IEEE},
    \uppercase{Sohei~Itahara}\authorrefmark{1}, \IEEEmembership{Student~Member,~IEEE},
    \uppercase{Kota~Yamashita}\authorrefmark{1}, \IEEEmembership{Student~Member,~IEEE},
    \uppercase{Koji~Yamamoto}\authorrefmark{1}, \IEEEmembership{Senior~Member,~IEEE},
    \uppercase{Yusuke~Koda}\authorrefmark{2}, \IEEEmembership{Member,~IEEE},
    \uppercase{Takayuki~Nishio}\authorrefmark{3}, \IEEEmembership{Senior~Member,~IEEE},
    \uppercase{and Akihito~Taya}\authorrefmark{4}, \IEEEmembership{Member,~IEEE},
}
\address[1]{Graduate School of Informatics, Kyoto University, Kyoto 606-8501, Japan (e-mail: kyamamot@i.kyoto-u.ac.jp)}
\address[2]{Centre of Wireless Communications, University of Oulu, 90014 Oulu, Finland (e-mail: Yusuke.Koda@oulu.fi)}
\address[3]{School of Engineering, Tokyo Institute of Technology, Ookayama, Meguro-ku, Tokyo, 152-8550, Japan (e-mail: nishio@ict.e.titech.ac.jp)}
\address[4]{Department of Integrated Information Technology, Aoyama Gakuin Univeristy, Fuchinobe, Chuo-ku, Sagamihara-shi, Kanagawa, 252-5258, Japan}

\tfootnote{
    This research and development work was supported in part by the MIC/SCOPE \#JP196000002 and JSPS KAKENHI Grant Number JP18H01442.
}

\markboth
{Author \headeretal: Preparation of Papers for IEEE TRANSACTIONS and JOURNALS}
{Author \headeretal: Preparation of Papers for IEEE TRANSACTIONS and JOURNALS}

\corresp{Corresponding author: Koji Yamamoto (e-mail: kyamamot@i.kyoto-u.ac.jp).}

\begin{abstract}
    In the field of WiFi sensing, as an alternative sensing source of the channel state information (CSI) matrix, the use of a beamforming feedback matrix (BFM) that is a right singular matrix of the CSI matrix has attracted significant interest owing to its wide availability regarding the underlying WiFi systems.
    In the IEEE 802.11ac/ax standard, the station (STA) transmits a BFM to an access point (AP), which uses the BFM for precoded multiple-input and multiple-output communications.
    In addition, in the same way, the AP transmits a BFM to the STA, and the STA uses the received BFM.
    Regarding BFM-based sensing, extensive real-world experiments were conducted as part of this study, and two key insights were reported:
    Firstly, this report identified a potential issue related to accuracy in existing uni-directional BFM-based sensing frameworks that leverage only BFMs transmitted for the AP or STA.
    Such uni-directionality introduces accuracy concerns when there is a sensing capability gap between the uni-directional BFMs for the AP and STA.
    Thus, this report experimentally evaluates the sensing ability disparity between the uni-directional BFMs,
    and shows that the BFMs transmitted for an AP achieve higher sensing accuracy compared to the BFMs transmitted from the STA
    when the sensing target values are estimated depending on the angle of departure of the AP.
    Secondly, to complement the sensing gap, this paper proposes a bi-directional sensing framework, which simultaneously leverages the BFMs transmitted from the AP and STA.
    The experimental evaluations reveal that bi-directional sensing achieves higher accuracy than uni-directional sensing in terms of the human localization task.
\end{abstract}


\titlepgskip=-15pt

\maketitle

\section{Introduction}
\label{sec:introduction}
WiFi sensing~\cite{yongsen2019wifi,zafari2019asurvey} has attracted notable interest as a technology that adds value to existing wireless local area networks (WLANs) beyond the communication infrastructure, which is under standardization by IEEE 802.11bf task group \cite{11bf}.
In WiFi sensing, a widely used radio frequency (RF) information is channel state information (CSI). It is used in multiple-input multiple-output orthogonal frequency-division multiplexing (MIMO-OFDM) systems~\cite{yongsen2019wifi}.
CSI is generally measured in the MIMO-OFDM communication procedures and includes high sensing capacity
to facilitate CSI-based sensing with low implementation cost and high sensing accuracy.

\begin{figure*}[t!]
    \centering
    \subfloat[Previous uni-directional sensing framework.]{%
        \includegraphics[width=0.45\textwidth, page =1]{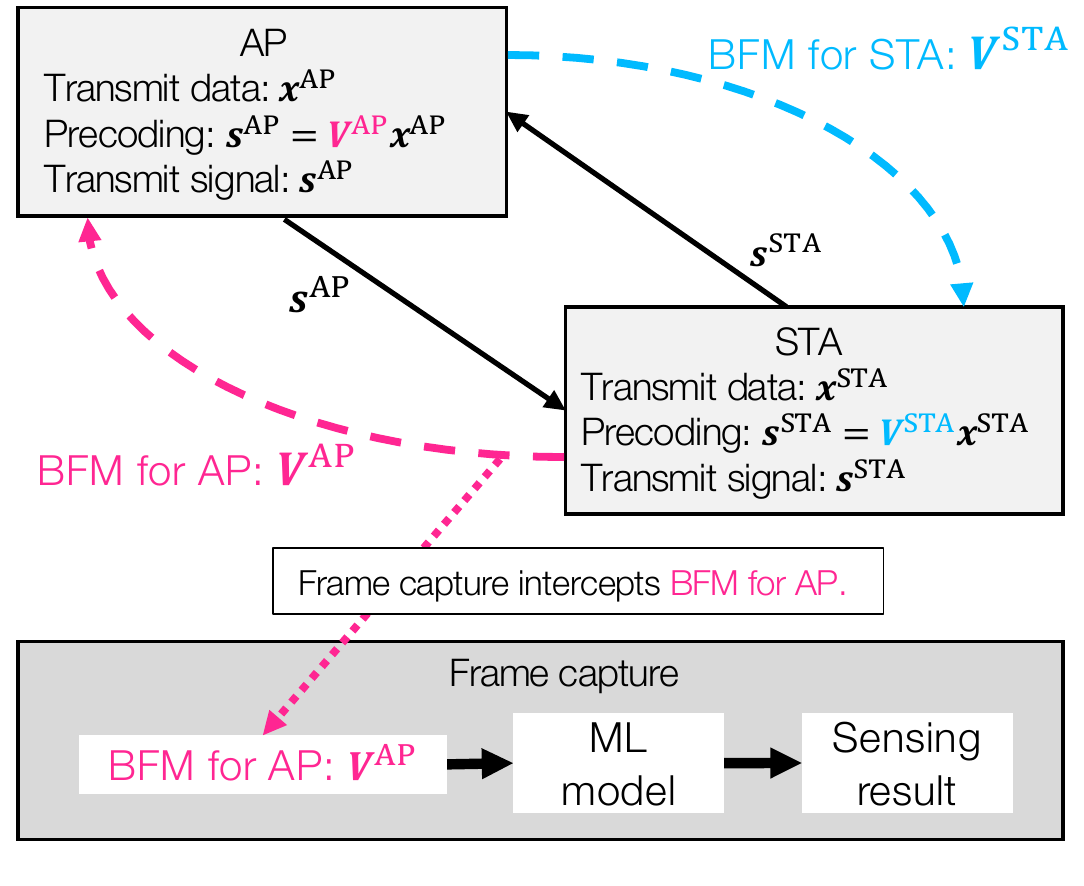}
    }
    \subfloat[\textit{Proposed bi-directional sensing framework.}]{%
        \includegraphics[width=0.45\textwidth, page =2]{img/ith_ver2.pdf}
    }
    \caption{
        Overview of previous and proposed BFM-based sensing frameworks.
        The previous framework uses only the uni-directional BFM (i.e., either $\bm{V}^\mathrm{AP}$ or $\bm{V}^\mathrm{STA}$) and ignores other directional BFM.
        The proposed framework uses the bi-directional BFMs (i.e., both of $\bm{V}^\mathrm{AP}$ and $\bm{V}^\mathrm{STA}$). }
    \label{fig:main}
\end{figure*}

CSI-based sensing is associated with an issue regarding the applicability of the underlying WLAN system.
Generally, access to the physical layer (PHY) component is necessary to obtain the CSI. However, only a few wireless chips permit such access to the PHY layer~\cite{gringoli2019free, csi_tool_11n_halperin,xie2015precise}.
Therefore, CSI-based sensing cannot necessarily be applied to most existing WLAN systems.
To extend their applicability, a new RF information, beamforming feedback matrix (BFM), has been utilized for sensing purposes~\cite{murakami2018wireless, miyazaki2019init, takahashi2019dnn,kanda2021respiratory,kato2021csi2image,fukushima2019evaluating}.
In the IEEE 802.11ac/ax standard~\cite{11ac, 11ax}, the BFM, which is a right singular matrix of the CSI matrix,
is transmitted from a station (STA) to an access point (AP) and is used for the precoding procedure in the AP for MIMO transmissions.
Moreover, in the same way, the AP transmits BFM to the STA, and the STA uses the received BFM for the precoding procedure in some scenarios, such as the STA that acts as a relay station.
The BFM transmission procedure is conducted without any encryption,
Thus, BFM-based sensing can be only conducted by capturing the BFM with a media-access-control (MAC) frame capture tool, without any access to the PHY layer of the communication pair.
This fact enables us to utilize most WLAN devices for BFM-based sensing.

We show an existing BFM-based sensing framework~\cite{murakami2018wireless, miyazaki2019init, takahashi2019dnn,kanda2021respiratory,kato2021csi2image,fukushima2019evaluating} in Fig.~\ref{fig:main}(a).
Let the BFM transmitted from the STA to the AP and the BFM transmitted from the AP to the STA be denoted by $\bm{V}^\mathrm{AP}$ and $\bm{V}^\mathrm{STA}$, respectively.
In these studies, the frame capture acquires BFMs and estimates the sensing target values (e.g., human locations~\cite{miyazaki2019init,murakami2018wireless, takahashi2019dnn}, device location~\cite{miyazaki2019init, takahashi2019dnn}, and respiratory rate~\cite{kanda2021respiratory}) by feeding BFMs to machine learning (ML) models.
The existing BFM-based sensing frameworks~\cite{murakami2018wireless, miyazaki2019init, takahashi2019dnn,kanda2021respiratory,kato2021csi2image} are referred to as uni-directional sensing, and they leverage either $\bm{V}^\mathrm{AP}$ or $\bm{V}^\mathrm{STA}$.
Therefore, even when the AP and STA transmit BFMs to each other, the existing works~\cite{murakami2018wireless, miyazaki2019init, takahashi2019dnn,kanda2021respiratory,kato2021csi2image} leverage only uni-directional BFMs (i.e., either of $\bm{V}^\mathrm{AP}$ or $\bm{V}^\mathrm{STA}$) and ignore the other directional BFMs.

Regarding the existing uni-directional sensing, we are concerned that there may be a sensing capability disparity between the usage of $\bm{V}^\mathrm{AP}$ and $\bm{V}^\mathrm{STA}$, resulting in the risk of using a BFM with a low sensing capability between $\bm{V}^\mathrm{AP}$ and $\bm{V}^\mathrm{STA}$.
The disparity is because $\bm{V}^\mathrm{AP}$ and $\bm{V}^\mathrm{STA}$ correspond to the right and left singular matrices of $\bm{H}$, respectively, and the right and left singular matrices of a matrix are generally different; thus, $\bm{V}^\mathrm{AP}$ and $\bm{V}^\mathrm{STA}$ are different.
This difference between $\bm{V}^\mathrm{AP}$ and $\bm{V}^\mathrm{STA}$ results in BFM disparity in the sensing accuracy.

To account for the accuracy disparity, we experimentally evaluate the sensing ability gap between $\bm{V}^\mathrm{AP}$ and $\bm{V}^\mathrm{STA}$ for the AP's AoD estimation task in a real environment using off-the-shelf equipment, which are equipped with non-linear antenna arrays.
The experimental evaluation confirmed that sensing using $\bm{V}^\mathrm{AP}$ resulted in a higher AoD estimation accuracy than sensing based on $\bm{V}^\mathrm{STA}$.
Moreover, this difference in the accuracy of AoD sensing implies that there is an accuracy difference for practical sensing tasks in which the sensing target values to be estimated depend on the AP's AoD.
Specifically, we experimentally evaluate the difference in accuracy between sensing with $\bm{V}^\mathrm{AP}$ and $\bm{V}^\mathrm{STA}$ using a human localization task in which the angle from the human to the AP corresponds to the AP's AoD of the human-reflected path.
The experimental results confirm the existence of a sensing accuracy disparity between the uni-directional BFMs.

Furthermore, in this report, a simple but powerful method called bi-directional sensing is proposed to address the potential accuracy concern.
An overview of the bi-directional sensing process is shown in Fig.~\ref{fig:main}(b).
In this method, the uni-directional BFMs are integrated into an input feature and are fed to the ML model.
Our experimental evaluations reveal that the proposed bi-directional sensing achieves higher sensing accuracy than the previous uni-directional sensing.
Moreover, it is determined that when the ML model is trained using the bi-directional BFMs, it leverages more appropriate BFM of the uni-directional BFMs.
Specifically, if the sensing with $\bm{V}^\mathrm{AP}$ achieves higher accuracy than sensing with $\bm{V}^\mathrm{STA}$,
the ML model with bi-directional BFMs assigns higher importance metrics to the input features generated from $\bm{V}^\mathrm{AP}$ compared to those of $\bm{V}^\mathrm{STA}$, and vice-versa.
Note that the importance metrics indicate the contribution of each input feature to the sensing accuracy.

The contributions of this study are summarized as follows:
\begin{enumerate}
    \item\label{cnt:bdb_sens_dif_aod}
          We experimentally validate that $\bm{V}^\mathrm{AP}$ achieves superior sensing accuracy than $\bm{V}^\mathrm{STA}$ for the AP's AoD estimation.
    \item \label{cnt:bdb_sens_dif_human}
          We experimentally validate the difference in the sensing accuracy between $\bm{V}^\mathrm{AP}$ and $\bm{V}^\mathrm{STA}$ for a human localization task, which is caused by the difference in sensing accuracy for the AP's AoD estimation.
          This finding highlights potential accuracy risks in existing BFM-based sensing schemes,
          which are not found in previous works that used only uni-directional BFM (i.e., either $\bm{V}^\mathrm{AP}$ or $\bm{V}^\mathrm{STA}$).
          To the best of our knowledge, in-depth discussions on the difference between uni-directional BFMs in terms of sensing accuracy have not been presented in the BFM-based sensing literature.
    \item \label{cnt:bdb_acc}
          We propose a novel BFM-based sensing framework called bi-directional sensing.
          In this approach, $\bm{V}^\mathrm{AP}$ and $\bm{V}^\mathrm{STA}$ are integrated into an input feature, which is fed to the ML model.
          We experimentally validate that the proposed bi-directional sensing achieved higher accuracy than the preexisting uni-directional sensing for a human localization task.
\end{enumerate}

In this study, our main objective is to show that the sensing abilities of the BFM transmitted for an AP and STA are different, and that the bi-directional BFM-based sensing framework is beneficial in terms of sensing accuracy when compared to uni-directional BFM-based sensing.
Namely, our focus is on the difference in the directivities of BFMs in BFM-based sensing frameworks.
Thus, the comparison of the proposed framework to other RF-information-based sensing frameworks (e.g., CSI-based sensing and received-power-based sensing) is out of the scope of this report.
Moreover, we should note that the BFM-based sensing framework is explicitly different from other RF-information-based sensing frameworks
in terms of its system requirements.
Specifically, the BFM-based sensing can be conducted using frame capture without access to the AP and STA, whereas the other RF-information-based sensing frameworks generally require such accessibility.

This study focuses on the difference in the sensing accuracy between three sensing methods: sensing with $\bm{V}^\mathrm{AP}$, $\bm{V}^\mathrm{STA}$, and both $\bm{V}^\mathrm{AP}$ and $\bm{V}^\mathrm{STA}$ (that is, $\bm{V}^\mathrm{AP}$ sensing, $\bm{V}^\mathrm{STA}$ sensing, and bi-directional sensing).
Thus, we consider that the evaluation of a scenario in which the training and testing datasets are generated in the same environment can be used to evaluate the difference.
It is beyond the scope of this study to provide a detailed evaluation of the train-test difference problem (that is, the problem that occurs when the environments of the training and testing datasets differ).

\section{Related Works}

\begin{table*}[t!]
    \caption{
        Summary of RF information used for WiFi sensing in terms of its system requirements.
    }
    \label{table:related_works_system}
    \centering
    \scalebox{1.}{
        \begin{tabular}{cccccccc}
            \toprule
            RF information & Firmware agnostic? & Need access to AP or STA? \\
            \midrule
            RSSI           & Yes                & Yes                       \\
            CSI            & No                 & Yes                       \\
            BFM            & Yes                & No                        \\
            \bottomrule
        \end{tabular}
    }
\end{table*}

\begin{table*}[t!]
    \caption{
        Summary of BFM-based WiFi sensing.
    }
    \label{table:related_works}
    \centering
    \scalebox{1.}{
        \begin{tabular}{cccccccc}
            \toprule
                                                                             & Task                                  & Need prior-training? & Bi-directional? \\
            \midrule
            \cite{miyazaki2019init,fukushima2019evaluating,takahashi2019dnn} & Human localization                    & Yes                  & No              \\
            \cite{itahara2021beamforming}                                    & AoD estimation                        & No                   & No              \\
            \cite{kanda2021respiratory}                                      & Respiratory rate estimation           & No                   & No              \\
            \cite{kato2021csi2image}                                         & Camera image reconstruction           & Yes                  & No              \\
            \textbf{Proposed method}                                         & Human localization and AoD estimation & Yes                  & Yes             \\
            \bottomrule
        \end{tabular}
    }
\end{table*}

Table~\ref{table:related_works_system} summarizes the system requirements of the existing WiFi sensing, by categorizing them into the received signal strength indicator RSSI-, CSI-, and BFM-based methods.
Traditionally, owing to its ease of availability and broad applicability, the received signal strength indicator (RSSI) has been used for WiFi sensing, such as human detection~\cite{mrazovac2012human}, human tracking~\cite{booranawong2019adaptive}, and human localization~\cite{booranawong2019implementation}.
Considering the spread of the MIMO system in WLAN, CSI-based sensing has attracted notable interest in terms of the improvement of the sensing capacity.
Since the CSI includes more fine-grained information than the RSSI, specifically CSI includes the attenuation between each transmit-receive antenna pair for each OFDM subcarrier, CSI-based sensing achieves higher sensing accuracy~\cite{qian2018enabling, qian2014pads, gu2017mosense, liu2017research} and success in more complex sensing tasks~\cite{chen2017rapid, cheng2016can, he2015wig, virmani2017position} than RSSI-based sensing.
In the existing CSI-based sensing literature, either of the firmwares~\cite{gringoli2019free,csi_tool_11n_halperin,xie2015precise} have been mainly used for CSI extraction.
However, they can only be used on a few wireless chips.
Therefore, there are device limitations in the realization of CSI-based sensing.

Table~\ref{table:related_works} summarizes the existing BFM-based sensing literature.
Compared to CSI-based sensing, BFM-based sensing is a firmware-agnostic wireless sensing method~\cite{miyazaki2019init, takahashi2019dnn,kanda2021respiratory,kato2021csi2image}.
As mentioned in the previous section, BFMs can be collected via MAC-layer frame capture without any special constraints regarding the firmware.
Although a vast number of studies addressed CSI-based sensing~\cite{yongsen2019wifi}, there are few studies on BFM-based sensing; human detection~\cite{murakami2018wireless, miyazaki2019init, takahashi2019dnn}, respiratory rate estimation\cite{kanda2021respiratory}, and camera image estimation~\cite{kato2021csi2image}.
Moreover, these experimental studies~\cite{murakami2018wireless,miyazaki2019init, takahashi2019dnn,kanda2021respiratory,kato2021csi2image} addressed sensing tasks using uni-directional BFM (i.e., either $\bm{V}^\mathrm{AP}$ or $\bm{V}^\mathrm{STA}$).
In contrast to those investigations ~\cite{murakami2018wireless,miyazaki2019init, takahashi2019dnn,kanda2021respiratory,kato2021csi2image},
this report focuses on the difference between the BFM transmitted for an AP and STA and leverages bi-directional BFMs to improve sensing accuracy.

\section{Preliminaries: MIMO-OFDM}
\label{sec:pre}
This section describes a MIMO-OFDM communication system using Eigen beam space division multiplexing (E-SDM)~\cite{miyashita2002high}.
The system consists of a transmitter (TX) and a receiver (RX) that are compliant with IEEE 802.11ac/11ax~\cite{11ac,11ax}.
The TX sends frames to the RX using MIMO-OFDM.
The RX estimates the CSI, computes the BFM based on the CSI, and transmits the BFM to the TX.
The TX uses the BFM as a precoding matrix.

Formally, let the CSI matrix from the TX to the RX at the $k$th subcarrier be denoted by $\bm{H}[k] \in \mathbb{C}^{N_{\mathrm{r}}\times N_{\mathrm{t}}}$,
where $N_\mathrm{t}$ and $N_\mathrm{r}$ are the number of antennas of the TX and RX, respectively.
The CSI matrix is estimated at the RX using the pilot signals (e.g., null data packet) at each OFDM subcarrier.
From the CSI matrix, the RX calculates a right singular matrix $\bm{V}[k]$ of $\bm{H}[k]$ using singular value decomposition, as
\begin{align}
    \label{eq:SVD}
    \bm{H}[k] = \bm{U}[k]\,\bm{\varSigma}[k]\,{\bm{V}[k]}^{\mathrm{H}},
\end{align}
where $\bm{V}[k]$ and $\bm{U}[k]$ are unitary matrices, and $\bm{\varSigma}[k]$ is a diagonal matrix with singular values.
Subsequently, the RX transmits the right singular matrix $\bm{V}[k]$, which is referred to as a BFM, to the TX using the BFM frame.
In the TX, the BFM is used for the precoding procedure.
Given a transmitting data vector $\bm{x}[k]$, the transmitted signal vector $\bm{s}[k]$ is denoted by
\begin{equation}
    \bm{s}[k] = \bm{V}[k]\,\bm{x}[k].
\end{equation}
In addition to $\bm{V}[k]$, the subcarrier-averaged substream gain $\bar{\bm{\varSigma}}$ is transmitted from the RX to the TX via the IEEE 802.11ac/11ax protocol \cite{11ac,11ax}, where
\begin{equation}
    \bar{\bm{\varSigma}} = \frac{1}{K}\sum_{k=1}^K \bm{\varSigma}[k],
\end{equation}
where $K$ is the number of subcarriers.

In the BFM transmission procedure of the IEEE 802.11ac/ax standards~\cite{11ac,11ax}, the BFM is quantized in the RX using the Givens transform to reduce the communication payload size of the BFM frame.
In this process~\cite{11ac,11ax}, the BFM $\bm{V}[k]$ is represented by an $M$-dimensional vector $\bm{v}^{'}[k] \in \mathbb{R}^M$, where $M$ is determined by $N_\mathrm{t}$ and $N_\mathrm{r}$ as follows:
\begin{align}
    M  & = 2N_\mathrm{t}N'-N'(N'+1),\label{equ:quant}        \\
    N' & \coloneqq \min(N_\mathrm{r}, N_\mathrm{t}-1).\notag
\end{align}

For shorthand notation, let the $M\times K$ matrix $\bm{V}^{'}$ denote the coordination of $(\bm{v}^{'}[k])^K_{k=1}$.
Moreover, the quantized BFM calculation function from the CSI matrices is denoted as $f^\mathrm{B}$,
where
\begin{align}
    \label{equ:BFM_cmp}
    \bm{V}^{'} = f^\mathrm{B}((\bm{H}[k])_{k=1}^{K}).
\end{align}
It should be noted that $\bm{V}^{'}$ represents information obtained via frame capture and is used for BFM-based sensing.\footnote{
    In this report, $\bm{V}[k]$ denotes the right singular matrix of the CSI matrix at the $k$th subcarrier, and $\bm{V}$ denotes the payload of the BFM matrix.
}

\section{Bi-directional Beamforming Feedback Matrix Sensing}
\label{sec:bi-directional_csi_feedback_meening}
\begin{figure}[t]
    \centering
    \subfloat[Training phase.]{%
        \includegraphics[width=0.45\linewidth, page =1]{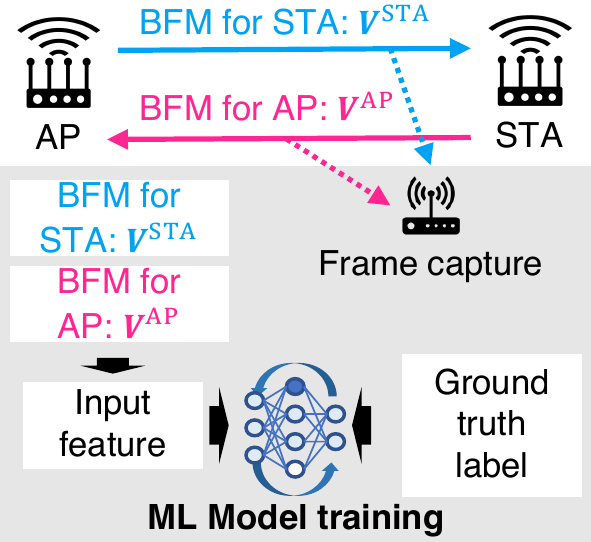}
    }
    \subfloat[Testing phase.]{%
        \includegraphics[width=0.45\linewidth, page =2]{img/ith_2.pdf}
    }
    \caption{Detailed procedure of bi-directional sensing frameworks.
        Note that frame capture facilitates WiFi sensing without any access to the STA and the AP.
    }
    \label{fig:detail}
\end{figure}

\subsection{System Model}
Fig.~\ref{fig:detail} shows the system model, which consists of an AP, an STA, and a frame capture device.
The AP and the STA periodically transmit MIMO frames between each other.
For the MIMO transmission, the BFM frames are transmitted from the AP to the STA, and from the STA to the AP over the air without encryption.
The frame capture obtains both the BFM transmitted from the AP and STA.
More formally, the CSI matrices from the AP to the STA and from the STA to the AP at the subcarrier $k$ are denoted as $\bm{H}^{\mathrm{AP}}[k]$ and $\bm{H}^{\mathrm{STA}}[k]$, respectively.
Based on Section~\ref{sec:pre}, the BFMs that are transmitted from the AP and STA are denoted as $\bm{V}^{\mathrm{AP}}$ and $\bm{V}^{\mathrm{STA}}$, respectively, where
\begin{align}
    \bm{V}^{\mathrm{AP}}  & = f^\mathrm{B}((\bm{H}^{\mathrm{AP}}[k])_{k=1}^{K}),  \\
    \bm{V}^{\mathrm{STA}} & = f^\mathrm{B}((\bm{H}^{\mathrm{STA}}[k])_{k=1}^{K}).
\end{align}

In this report, based on existing BFM-based sensing methods, an ML-based sensing technique is developed.
Thus, the system has two-time phases: a training phase and a testing phase.
In the training phase, the frame capture obtains BFMs and the ground-truth target label (e.g., actual measured location of a human subject), and the BFMs are used as input features.
The ML model is trained using a tranining dataset consiting of the input features and target labels.
In the testing phase, whenever the frame capture obtains the BFM frame, it estimates the target label by feeding the BFM to the trained ML model.
Additionally, as with the existing sensing with prior training, the system model in this report requires rebuilding the ML model when the input dimension of the ML model is changed (for example, when the number of antennas of the AP and STA are changed).

\subsection{BFM disparity}
\textbf{(2-1, 2-2, and 2-8)}
The disparity between the two BFMs in sensing accuracy (that is, a difference in the sensing accuracies of $\bm{V}^\mathrm{AP}$ and $\bm{V}^\mathrm{STA}$) is suspected for two reasons.
First, if there exists channel reciprocity between the AP and STA, there is generally no BFM reciprocity (that is, $\bm{V}^\mathrm{AP}$ generally differs from $\bm{V}^\mathrm{STA}$), which is detailed in the following paragraph.
\textbf{(1-8)}
Second, although the AoD of the AP can be estimated from $\bm{V}^\mathrm{AP}$~\cite{itahara2021beamforming}, the sensing accuracy of the AoD of the AP using $\bm{V}^\mathrm{STA}$ and bi-directional BFMs (that is, using both $\bm{V}^\mathrm{STA}$ and $\bm{V}^\mathrm{AP}$) is not clear.

Even if channel reciprocity exists between the AP and the STA, BFM reciprocity does not exist.
Specifically, even if channel reciprocity exists, $\bm{V}^\mathrm{AP}$ differs from $\bm{V}^\mathrm{STA}$,
because $\bm{V}^\mathrm{AP}$ and $\bm{V}^\mathrm{STA}$ correspond to right and left singular matrices of $\bm{H}$, respectively, and
generally, the right and left singular matrices of a matrix are different.
Thus, $\bm{V}^\mathrm{AP}$ and $\bm{V}^\mathrm{STA}$ are different, and
the difference between $\bm{V}^\mathrm{AP}$ and $\bm{V}^\mathrm{STA}$ results in BFM disparity in the sensing accuracy.

\renewcommand{\thefootnote}{\arabic{footnote}}
Specifically, given the channel reciprocity, the CSI matrix from the STA to the AP (that is, $\bm{H}^\mathrm{STA}$) is represented by
\begin{align}
    \bm{H}^\mathrm{STA} = (\bm{H}^\mathrm{AP})^\mathrm{T}.\label{equ:r1}
\end{align}
Further, $\bm{V}^\mathrm{AP}$ and $\bm{V}^\mathrm{STA}$ are right singular matrices of $\bm{H}^\mathrm{AP}$ and $\bm{H}^\mathrm{STA}$, respectively.
\begin{align}
    \bm{H}^\mathrm{STA} & = \bm{U}^\mathrm{STA}\bm{\varSigma}^\mathrm{STA}(\bm{V}^\mathrm{STA})^\mathrm{H} \label{equ:r2} \\
    \bm{H}^\mathrm{AP}  & = \bm{U}^\mathrm{AP}\bm{\varSigma}^\mathrm{AP}(\bm{V}^\mathrm{AP})^\mathrm{H}.\label{equ:r3}
\end{align}
Substituting \eqref{equ:r1} to \eqref{equ:r2},
\begin{align}
    (\bm{H}^\mathrm{AP})^\mathrm{T} & = \bm{U}^\mathrm{STA}\bm{\varSigma}^\mathrm{STA}(\bm{V}^\mathrm{STA})^\mathrm{H} \notag                             \\
    \bm{H}^\mathrm{AP}              & = (\bm{V}^\mathrm{STA})^{*}(\bm{\varSigma}^\mathrm{STA})^\mathrm{T}(\bm{U}^\mathrm{STA})^\mathrm{T}.\label{equ:r4}
\end{align}
From \eqref{equ:r4}, $\bm{V}^\mathrm{STA}$ corresponds to a left singular matrix of $\bm{H}^\mathrm{AP}$.
By comparing \eqref{equ:r3} and \eqref{equ:r4}, $\bm{V}^\mathrm{AP}$ and $\bm{V}^\mathrm{STA}$ correspond to right and left singular matrices of $\bm{H}^\mathrm{AP}$, respectively,
\footnote{
    Although $\bm{U}^\mathrm{AP}$ and $(\bm{V}^\mathrm{STA})^{*}$ correspond to the right singular matrix of $\bm{H}^\mathrm{AP}$,
    $\bm{U}^\mathrm{AP}$ and $(\bm{V}^\mathrm{STA})^{*}$ are not necessarily the same.
    This is because provided an arbitrary matrix $\bm{H}$, multiple matrices can be its right singular matrix.
}.
Generally, a left and right singular matrix of a particular matrix are independent.
Thus, $\bm{V}^\mathrm{AP}$ differs from $\bm{V}^\mathrm{STA}$, resulting in a BFM disparity in the sensing accuracy.

\section{Experimental Setup}
\label{sec:exp_setup}
We experimentally evaluated the accuracy of BFM-based sensing methods for two sensing tasks, AoD estimation and human localization, using off-the-shelf WiFi devices in outdoor and indoor environments.
These sensing tasks are formulated as classification problems.
For shorthand notation, we denote $\bm{V}^\mathrm{AP}$ sensing and $\bm{V}^\mathrm{STA}$ sensing as uni-directional sensing with $\bm{V}^\mathrm{AP}$ and $\bm{V}^\mathrm{STA}$, respectively.

\subsection{System Components}
\begin{figure}[t!]
    \centering
    \includegraphics[width=\hsize]{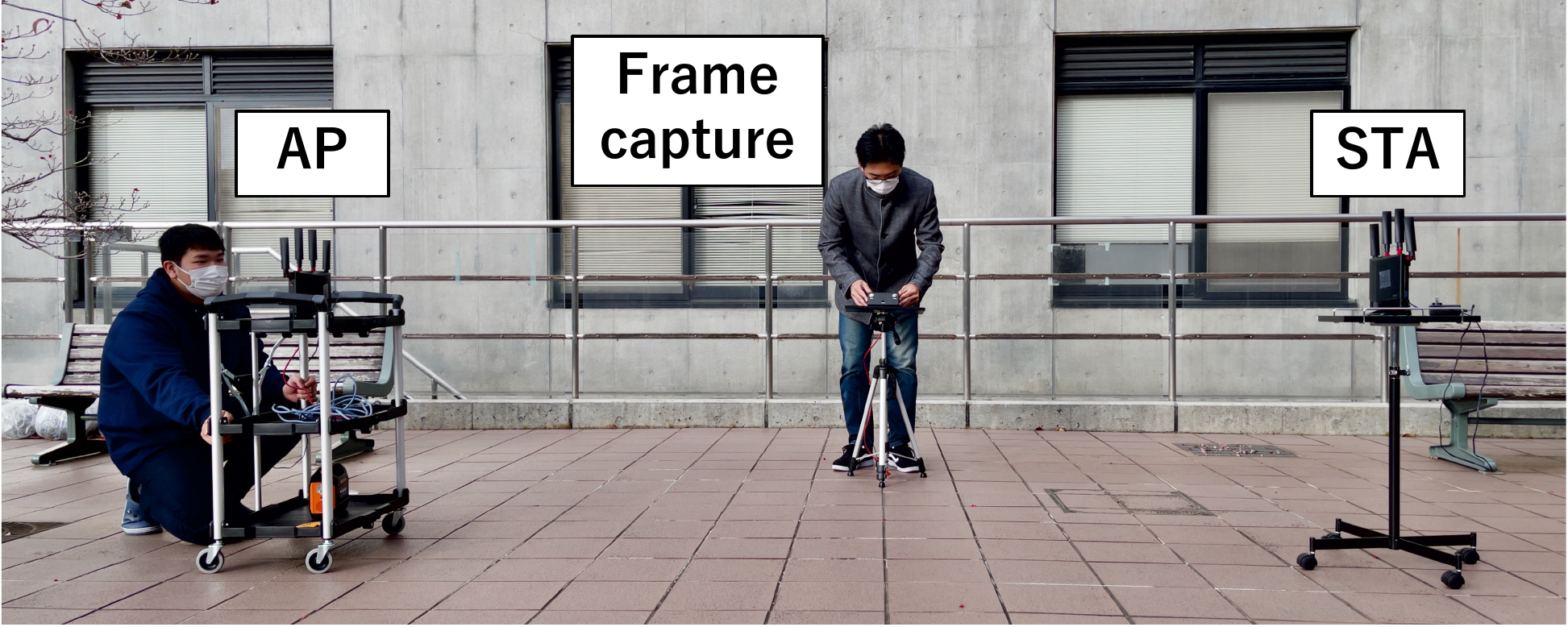}
    \caption{Layout of experimental setup.
        AP, STA, and frame capture are at a height of 75 cm}
    \label{ex_actual}
\end{figure}

As depicted in Fig.~\ref{ex_actual}, the system consists of an AP, an STA, and frame capture.
The AP and STA are equipped with four antennas.
Since the number of antennas of the AP and STA is identical, the dimension of the BFM at each subcarrier is the same among $\bm{V}^\mathrm{AP}$ and $\bm{V}^\mathrm{STA}$, and $M$ is 12.\footnote{The dimension $M$ is determined by the number of the antenna of the AP and STA following~\eqref{equ:quant}.}
In this evaluation, the quantized bit widths $\bm{V}^\mathrm{AP}$ and $\bm{V}^\mathrm{STA}$ were the same and followed the IEEE 802.11ac~\cite{11ac} standard.

This study evaluated BFM-based sensing using either of two sets of equipment: equipment set A and equipment set B, which are listed in Table~\ref{table:equipment}.
For both equipment sets, the same products were used for the AP and STA (that is, the chipset and antenna array were identical among the AP and STA).
Equipment set A and B comply with IEEE 802.11ax and IEEE 802.11ac, respectively.
For equipment set A and B, the number of subcarriers $K$ were 64 and 52, respectively, resulting in a BFM of $12\times 64$ and $12\times 52$, respectively.
Moreover, all the equipment was off-the-shelf devices.

We loaded heavy traffic using iperf in both uplink and downlink.
Specifically, the throughput of the uplink and downlink were set as 100\,Mbit/s so that the AP and STA transmit BFMs at an average interval of 0.1\,s.
Note that, in this evaluation, the AP and STA are connected such that the STA acts as a relay station.

\begin{table}[t!]
    \caption{Experimental equipment.}
    \label{table:equipment}
    \centering
    \subfloat[][Equipment set-A.]{
        \scalebox{0.90}[0.90] {
            \renewcommand\arraystretch{1.1}
            \begin{tabular}{cc}
                \toprule
                AP and STA                        & Buffalo WXR-5700AX7S \\
                Wireless chipset of AP and STA    & BCM4910              \\
                Frame capture                     & NVIDA Jetson nano    \\
                Wireless chipset of frame capture & Intel AX200          \\
                Protocol                          & IEEE 802.11ax        \\
                Wireless band                     & 104\,ch              \\
                Bandwidth                         & 20\,MHz              \\
                \bottomrule
            \end{tabular}
        }
    }\\
    \subfloat[][Equipment set B.]{
        \scalebox{0.90}[0.90] {
            \renewcommand\arraystretch{1.1}
            \begin{tabular}{cc}
                \toprule
                AP and STA                        & ASUS RT-AC66U     \\
                Wireless chipset of AP and STA    & BCM4706           \\
                Frame capture                     & NVIDA Jetson nano \\
                Wireless chipset of frame capture & Intel AX200       \\
                Protocol                          & IEEE 802.11ac     \\
                Wireless band                     & 104\,ch           \\
                Bandwidth                         & 20\,MHz           \\
                \bottomrule
            \end{tabular}
        }
    }
\end{table}

\subsection{Experimental Scenario}
The experimental evaluation uses two sensing tasks: AoD estimation and human localization, in two real-world environments: an outdoor and an indoor environments, respectively.
Unless otherwise noted, the evaluation was conducted with equipment set A.
The evaluation of the human localization task in the outdoor environment was conducted using either equipment set A or equipment set B.
It should be noted that the evaluation aims to compare the three sensing methods using the same equipment in the same environment.
Thus, we avoided comparing the results obtained from different environments or equipment.

\textbf{AoD estimation.}
This evaluation aims to assess the accuracy difference of the AP's AoD estimation on the realistic environment of two uni-directional sensing approaches: sensing using $\bm{V}^\mathrm{AP}$ and sensing using $\bm{V}^\mathrm{STA}$.
Specifically, we estimate the AoD of the line-of-sight path.

\begin{figure}[t!]
    \centering
    \subfloat[Angle of departure estimation.]{%
        \includegraphics[width=0.47\hsize]{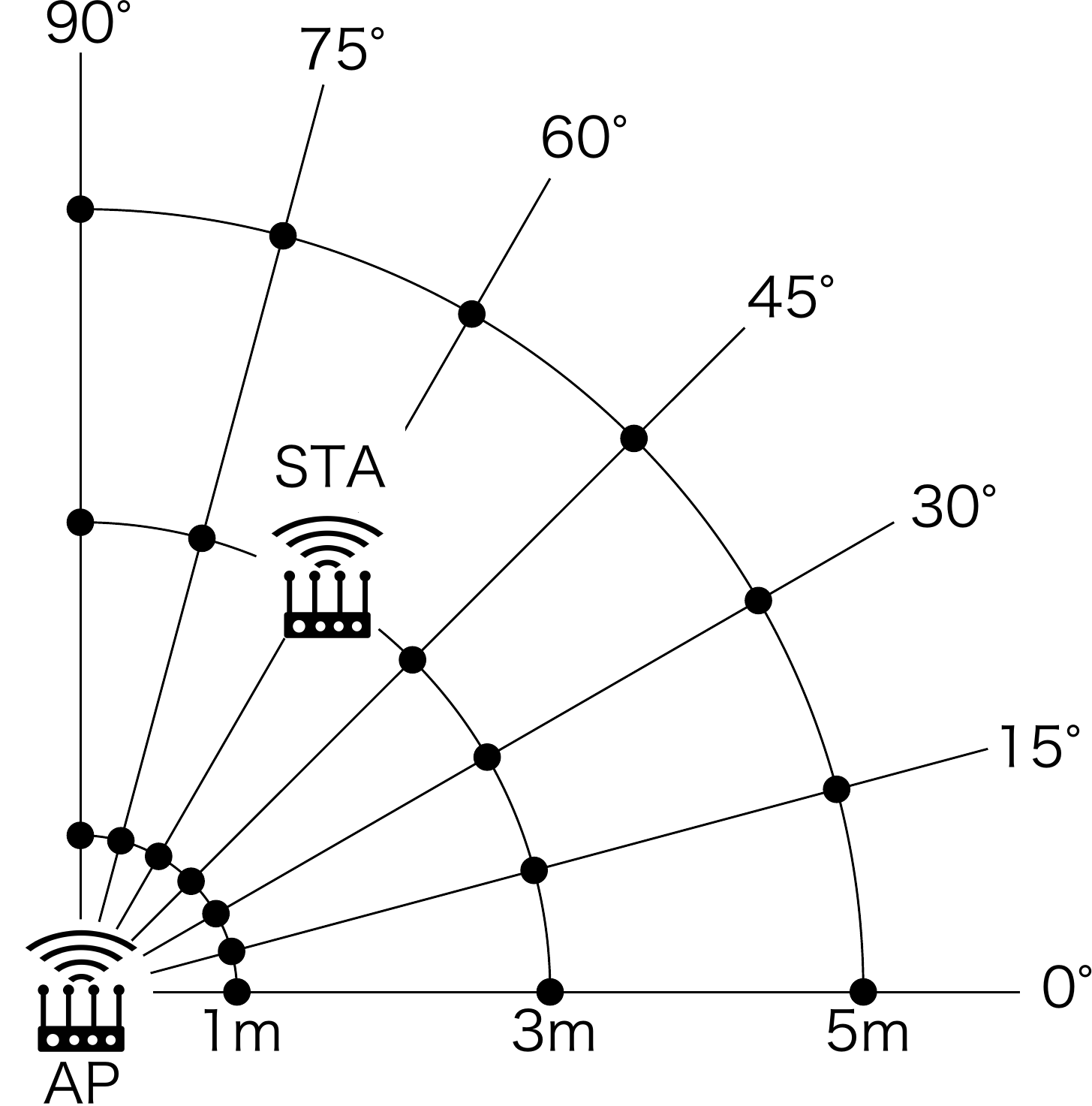}
    }
    \subfloat[Human localization.]{%
        \includegraphics[width=0.47\hsize]{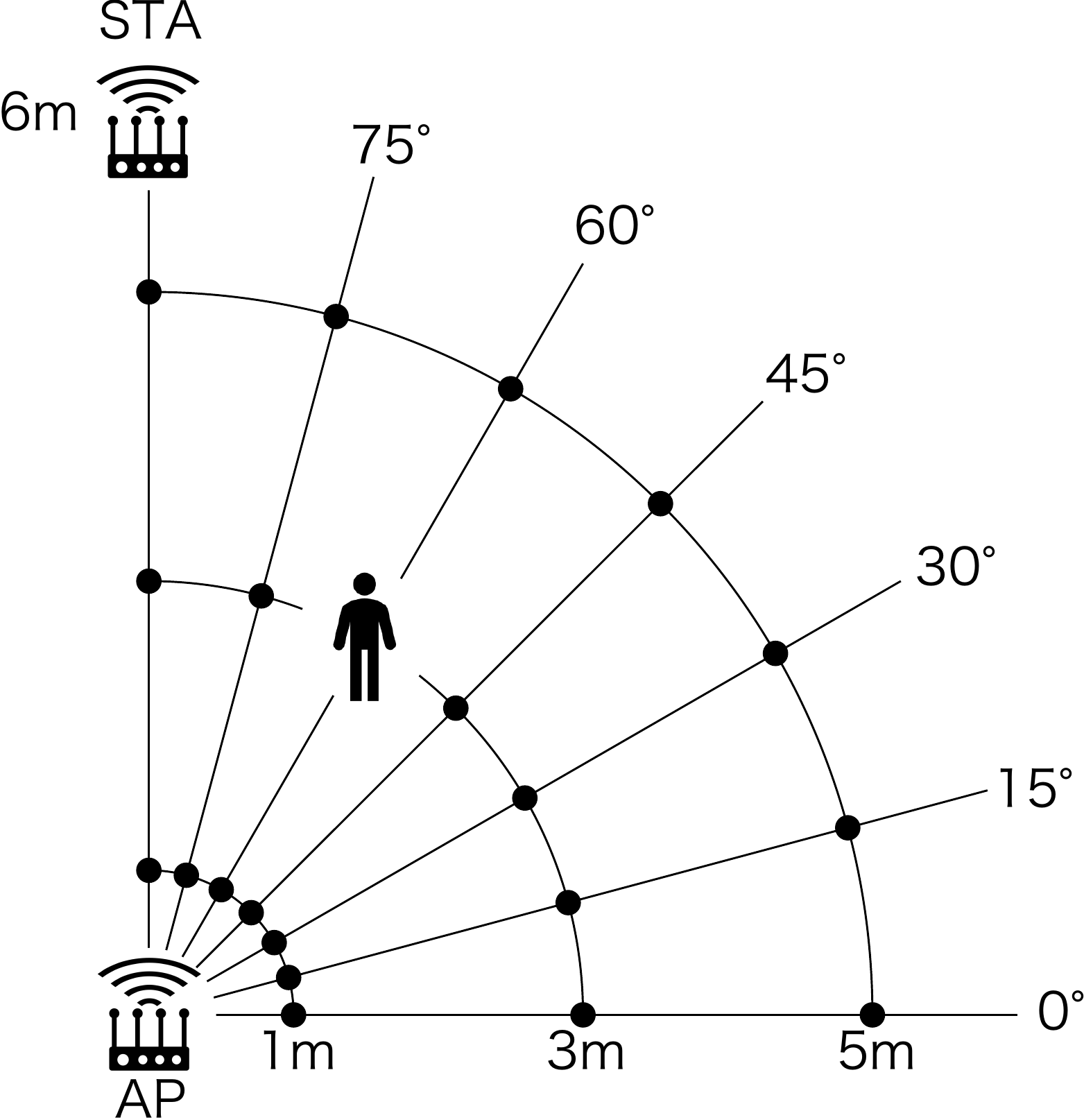}
    }
    \caption{
        Equipment deployment in outdoor environment.
        Preparation of a polar coordinate system centered on the AP.
        The antenna array of the AP is placed parallel to the zero-degree direction.
        For the AoD estimation task, the STA is located at any of 21 points, and is depicted as black dots in Fig.~\ref{fig:experimental_environments}(a).
        For the human localization task, the STA is fixed at the position $(6\,\mathrm{m}, 90\tcdegree)$, while
        a human stands at any of the 21 points and is depicted as black dots in Fig.~\ref{fig:experimental_environments}(b).
    }
    \label{fig:experimental_environments}
\end{figure}

\begin{figure}[t!]
    \centering
    \subfloat[Angle of departure estimation.]{%
        \includegraphics[width=0.47\hsize, page =1]{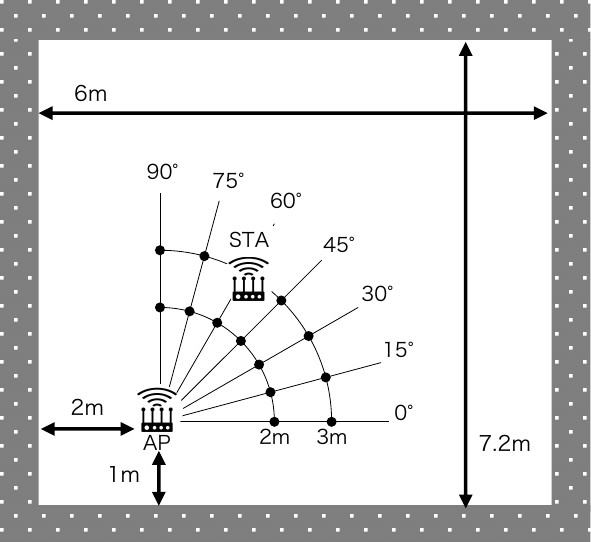}
    }
    \subfloat[Human localization.]{%
        \includegraphics[width=0.47\hsize, page =2]{img/indoor_kondo.pdf}
    }
    \caption{
        Equipment deployment in indoor environment.
        The AP and human are located on either of the 14 points depicted by the black dots for AoD estimation and human localization, respectively.
    }
    \label{fig:experimental_environments_indoor}
\end{figure}

Figs.~\ref{fig:experimental_environments}(a) and \ref{fig:experimental_environments_indoor}(a) show the outdoor and indoor environment, respectively, where we generated a dataset consisting of seven classes in terms of AoD,
which is either of seven angles $\{0\tcdegree, 15\tcdegree, 30\tcdegree, 45\tcdegree, 60\tcdegree, 75\tcdegree, 90\tcdegree\}$.
In the outdoor and indoor environments, in each class, the STA is located at one of three and two positions, respectively.
Specifically, the distances between each position and the AP are given as
$\{1\,\mathrm{m}, 3\,\mathrm{m}, 5\,\mathrm{m}\}$ and $\{2\,\mathrm{m}, 3\,\mathrm{m}\}$, respectively.
The AoD only depends on the position of the STA.
In the outdoor and indoor experiments, we obtained 12,600 and 1,800 data samples, of which 1,800 and 200 samples corresponded to each AoD, respectively.
The orientation of the antenna array of the STA was randomly changed throughout the experiment.
The AoD depended only on the position of the STA and not on the orientation of the antenna array of the STA.

\textbf{Human localization.}
This evaluation aims to assess the accuracy of the uni-directional sensing and the proposed bi-directional sensing approaches on more practical sensing tasks than AoD estimation.
Figs.~\ref{fig:experimental_environments}(b) and \ref{fig:experimental_environments_indoor}(b) show the overviews of the outdoor and indoor environments, respectively.
We generated a dataset wherein a human was located at any of the 21 and 14 positions in the outdoor and indoor environments, respectively.
The positions are denoted using the distance $r$ and the angle $\theta$ to the AP.
As such, the target label is represented by a two-dimensional vector $(r, \theta)$.
In this scenario, two ML models are trained to estimate the angle $\theta$ and the distance $d$.
The positions of the STA are fixed.
It should be noted that $\theta$ corresponds to the AP's AoD of the human-reflected path in this experimental scenario.

This evaluation was conducted using either experimental equipment set A or equipment set B.
When using equipment set A, we obtained 12,600 and 1,600 data samples, wherein 600 and 200 samples corresponded to each position in the outdoor and indoor environments, respectively.
In the case of equipment set B, we obtained 4,200 data samples, of which 200 samples corresponded to each position in the outdoor environment.

\subsection{Machine Learning}
Three ML models are utilized: a random forest (RandF)~\cite{breiman2001random}, a light gradient boosting machine (LightGBM)~\cite{ke2017lightgbm}, and support vector machine (SVM)~\cite{cortes1995support}.
The AoD estimation and the human localization are formulated as the classification problem.
In this evaluation, the dataset is randomly divided into training and testing datasets with a ratio of 9:1.
When using the RandF and LightGBM, we performed 10-fold leave-one-out cross-validation for ten trials with a different random seed.
When using the SVM, we did not conduct cross-validation and the ML model was trained only once.

The hyperparameters are selected as follows unless otherwise indicated.
For the RandF, the maximum depth, the splitting criterion, and the number of trees are selected as 5, Gini impurity, and 50, respectively.
For the LightGBM, the maximum depth, the splitting criterion, the number of trees, and the learning rate are selected as infinite, multi-class log loss, 5, and 0.1, respectively.
For the SVM, the regularization parameter and the kernel are selected as 1.0, and the Gaussian kernel, respectively.

\subsection{Feature Generation}
In bi-directional sensing, the bi-directional BFMs are integrated to generate an input feature.
For equipment set A and B, we used a different method to generate bi-directional BFMs.
When using equipment-set-A, given that $\bm{V}^\mathrm{AP}$ and $\bm{V}^\mathrm{STA}$ are captured within a time interval of less than $t_0$, they are flattened and concatenated.
The input feature vector with a dimension of 1,536 is then generated.
In the experimental evaluation process, $t_0$ is set to 0.15\,s.

When using equipment set B, for each target class (that is, human location or AoD of the AP),
we first obtained $\bm{V}^\mathrm{AP}$ and subsequently, obtained $\bm{V}^\mathrm{STA}$;
then $\bm{V}^\mathrm{AP}$ and $\bm{V}^\mathrm{STA}$ were randomly integrated into an input feature of bi-directional sensing.
The input feature vector with a dimension of 1,248 was then generated.

However, in uni-directional sensing, either $\bm{V}^\mathrm{AP}$ or $\bm{V}^\mathrm{STA}$ is used.
To allow for a fair comparison between uni-directional sensing and bi-directional sensing,
the former uses the input feature, for which the dimension is the same as that of bi-directional sensing.
Thus, two BFMs that were captured within a time interval of less than $t_0$ are flattened and concatenated, and the input feature vector is then generated.

\section{Result}
\subsection{Angle of Departure Estimation}
\label{ssc:acc_rate_cmp}

\begin{table}[t!]
    \centering
    \caption{
        Classification accuracy of seven classes and average error of AoD estimation using three ML models.
        $\bm{V}^\mathrm{AP}$ sensing achieved higher AP AoD accuracy than $\bm{V}^\mathrm{STA}$ sensing.
    }
    \label{tab:devicelocalization_acc_summary}

    \centering
    \subfloat[][Classification accuracy. ]{
        \begin{tabular}{ccccccc}
            \toprule
                                     & ML model & $\bm{V}^\mathrm{AP}$ sensing & $\bm{V}^\mathrm{STA}$  sensing \\
            \midrule
            \multirow{3}{*}{Outdoor} & RandF    & \textbf{98.7}\%              & 56.6\%                         \\
                                     & LightGBM & \textbf{99.7}\%              & 78.0\%                         \\
                                     & SVM      & \textbf{99.9}\%              & 87.0\%                         \\
            \midrule
            \multirow{3}{*}{Indoor}  & RandF    & \textbf{99.2}\%              & 81.5\%                         \\
                                     & LightGBM & \textbf{93.2}\%              & 92.0\%                         \\
                                     & SVM      & \textbf{99.3}\%              & 92.9\%                         \\
            \bottomrule
        \end{tabular}}
    \\
    \subfloat[][Average error.]{
        \begin{tabular}{cccc}
            \toprule
                                     & ML model & $\bm{V}^\mathrm{AP}$ sensing & $\bm{V}^\mathrm{STA}$  sensing \\
            \midrule
            \multirow{3}{*}{Outdoor} & RandF    & \textbf{0.21}\textdegree     & 14.5\textdegree                \\
                                     & LightGBM & \textbf{0.09}\textdegree     & 6.58\textdegree                \\
                                     & SVM      & \textbf{0.02}\textdegree     & 4.46\textdegree                \\
            \midrule
            \multirow{3}{*}{Indoor}  & RandF    & \textbf{0.89}\textdegree     & 5.61\textdegree                \\
                                     & LightGBM & \textbf{0.88}\textdegree     & 3.79\textdegree                \\
                                     & SVM      & \textbf{0.21}\textdegree     & 2.79\textdegree                \\
            \bottomrule
        \end{tabular}
    }
\end{table}

In this section, the results show that higher accuracy was obtained for $\bm{V}^\mathrm{AP}$ sensing in the process of the AoD estimation of the AP than for $\bm{V}^\mathrm{STA}$ sensing, which validates contribution~\ref{cnt:bdb_sens_dif_aod} in section~\ref{sec:introduction}.
As shown in Table~\ref{tab:devicelocalization_acc_summary}(a), regardless of the ML model and the experimental environment,
$\bm{V}^\mathrm{AP}$ sensing achieved higher accuracy than $\bm{V}^\mathrm{STA}$ sensing in the AoD estimation of the AP.
Moreover, in an outdoor environment, the accuracy of $\bm{V}^\mathrm{AP}$ sensing was higher than 0.98 for the three ML models, indicating that the performance was almost perfect.
Table~\ref{tab:devicelocalization_acc_summary}(b) shows the average error for AoD estimation using the three ML models.
The average error for $\bm{V}^\mathrm{AP}$ sensing was much smaller than that for $\bm{V}^\mathrm{STA}$ sensing, regardless of the ML model used.
Specifically, regardless of the ML model, the average error for $\bm{V}^\mathrm{AP}$ sensing was lower than 0.3\textdegree and 0.9\textdegree in outdoor and indoor environments, respectively.
However, the error of $\bm{V}^\mathrm{STA}$ sensing was larger than 4.0\textdegree and 2.5\textdegree in outdoor and indoor environments, respectively.

\begin{figure}[t!]
    \centering
    \includegraphics[width=0.45\textwidth]{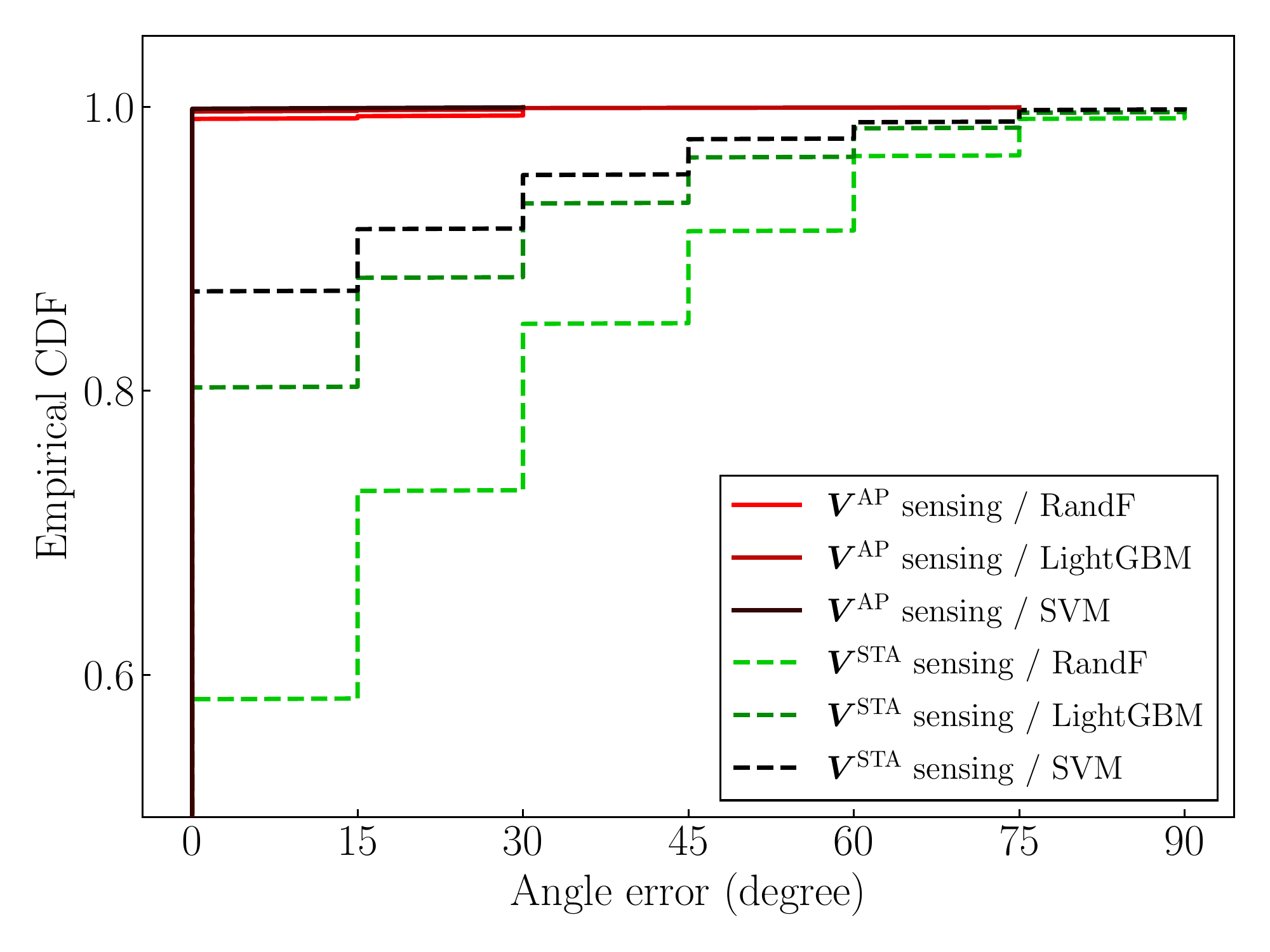}
    \caption{
        Empirical CDF of estimation error in AoD estimation using three ML models.
        The red and green lines represent the results for $\bm{V}^\mathrm{AP}$ and $\bm{V}^\mathrm{STA}$ sensing, respectively.
    }
    \label{fig:devicelocalization_error_hist_angle}
\end{figure}

\begin{figure}[t!]
    \centering
    \includegraphics[width=0.4\textwidth]{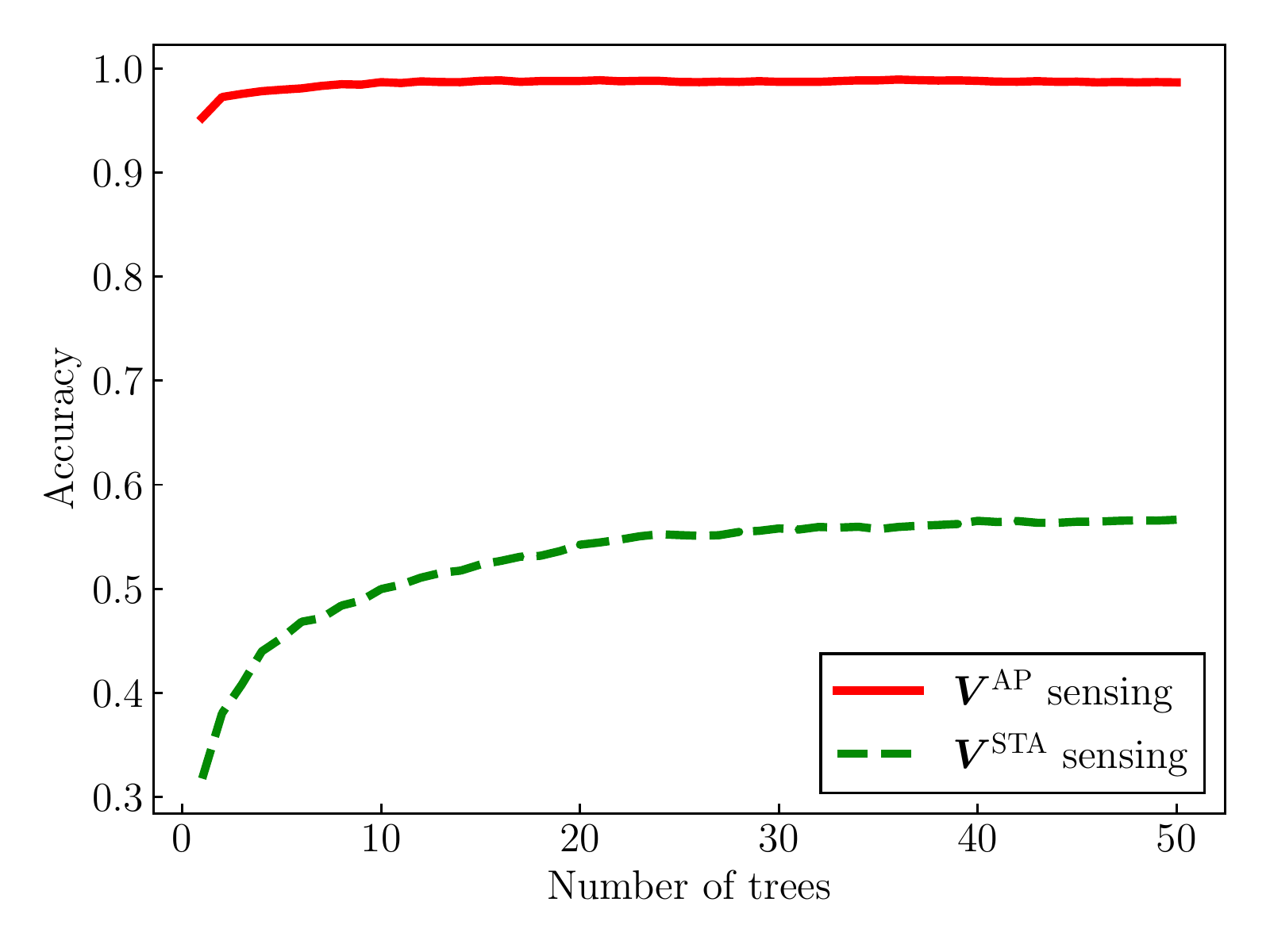}
    \caption{
        Effect of the number of trees of RandF on AoD estimation accuracy.
    }
    \label{fig:devicelocalization_acc}
\end{figure}

Fig.~\ref{fig:devicelocalization_error_hist_angle} shows the empirical cumulative distribution function (CDF) of the AoD estimation error in the outdoor environment.
Regardless of the ML model, in the case of $\bm{V}^\mathrm{AP}$ sensing, more than 99\% of the test samples had an error less than 30\textdegree, whereas for $\bm{V}^\mathrm{STA}$ sensing, less than 92\% of the samples met this criterion. In addition, the effect of the ML hyperparameters on accuracy in the RandF model is shown in Fig.~\ref{fig:devicelocalization_acc}.
This finding is consistent with the results described so far; the accuracy for $\bm{V}^\mathrm{AP}$ sensing is higher than that of $\bm{V}^\mathrm{STA}$ sensing, regardless of the number of trees.
Thus, we can conclude that $\bm{V}^\mathrm{AP}$ sensing achieves higher AP AoD sensing accuracy compared to $\bm{V}^\mathrm{STA}$ sensing.

\subsection{Human Localization}
\label{ssc:error_dist_comp}
\noindent\textbf{Accuracy comparison.}
In this section, the accuracy of the three BFM-based sensing methods (i.e., $\bm{V}^\mathrm{AP}$ sensing, $\bm{V}^\mathrm{STA}$ sensing, and bi-directional BFM-based sensing) was evaluated based on a human localization task.
These parameters were considered as part of the evaluation including the angle, distance, and position.

Table~\ref{tab:humanlocalization_acc_summary} shows the three accuracy metrics for the three BFM-based sensing methods.
In the case of angle accuracy, which is shown in Table~\ref{tab:humanlocalization_acc_summary}(a), $\bm{V}^\mathrm{AP}$ sensing achieved higher accuracy than $\bm{V}^\mathrm{STA}$ sensing, regardless of the ML model.
Considering that the angle $\theta$ corresponds to the AoD of the AP of the human-reflected path in the experimental setup,
the difference in angle accuracy is because $\bm{V}^\mathrm{AP}$ includes more useful information for the AoD of the AP than the BFM $\bm{V}^\mathrm{STA}$, as indicated in section~\ref{ssc:acc_rate_cmp}.
Owing to the difference in angle accuracy, $\bm{V}^\mathrm{AP}$ sensing achieved a higher position accuracy than $\bm{V}^\mathrm{STA}$ sensing, as shown in Table~\ref{tab:humanlocalization_acc_summary}(c).
Thus, we can conclude that there is a difference in the sensing capabilities of $\bm{V}^\mathrm{AP}$ and $\bm{V}^\mathrm{STA}$ in terms of the human localization task, which validates contribution~\ref{cnt:bdb_sens_dif_human} in section~\ref{sec:introduction}.
However, in terms of the distance accuracy, as shown in Table~\ref{tab:humanlocalization_acc_summary}(b),
the accuracy of $\bm{V}^\mathrm{STA}$ sensing was comparable to that of $\bm{V}^\mathrm{AP}$ sensing, regardless of the ML model.
This implies that the variability of $\bm{V}^\mathrm{STA}$ in terms of human-distance estimation is comparable to that of $\bm{V}^\mathrm{AP}$.

As shown in Table~\ref{tab:humanlocalization_acc_summary}, bi-directional sensing achieved higher accuracy compared to uni-directional sensing in terms of the accuracy metrics and ML models, which validates contribution \ref{cnt:bdb_acc} in section~\ref{sec:introduction}.
This difference in accuracy is because the ML model that is trained based on bi-directional BFMs leverages the more appropriate BFM of the two uni-directional BFMs, which is validated in the following section.
The difference in accuracy between bi-directional and uni-directional sensing is more robustly validated in terms of the localization error in the following section.

\begin{table}[t]
    \centering
    \caption{
        Classification accuracy of human localization with equipment set A.
        For the outdoor environment, the angle, distance, and position accuracy were defined as the classification accuracy of 7, 3, and 21 classes, respectively.
        For the indoor environment, they were defined as the classification accuracy of 7, 2, and 14 classes, respectively.
    }
    \label{tab:humanlocalization_acc_summary}
    \subfloat[][Angle accuracy: $\theta$.]{
        \begin{tabular}{ccccc}
            \toprule
                                     & ML model & Bi-directional & $\bm{V}^\mathrm{AP}$ & $\bm{V}^\mathrm{STA}$ \\

            \midrule
            \multirow{3}{*}{Outdoor} & RandF    & \bf{87.3}\%    & 82.4\%               & 71.8\%                \\
                                     & LightGBM & \bf{94.8}\%    & 93.4\%               & 86.6\%                \\
                                     & SVM      & \bf{98.5}\%    & 96.6\%               & 95.2\%                \\
            \midrule
            \multirow{3}{*}{Indoor}  & RandF    & \bf{95.5}\%    & 92.2\%               & 90.5\%                \\
                                     & LightGBM & \bf{94.5}\%    & 94.5\%               & 93.4\%                \\
                                     & SVM      & \bf{98.6}\%    & 96.0\%               & 92.9\%                \\
            \bottomrule
        \end{tabular}
    }\\
    \subfloat[][Distance accuracy: $d$.]{
        \begin{tabular}{ccccc}
            \toprule
                                     & ML model & Bi-directional & $\bm{V}^\mathrm{AP}$ & $\bm{V}^\mathrm{STA}$ \\
            \midrule
            \multirow{3}{*}{Outdoor} & RandF    & \bf{83.6}\%    & 78.4\%               & 78.4\%                \\
                                     & LightGBM & \bf{90.6}\%    & 86.2\%               & 88.1\%                \\
                                     & SVM      & \bf{96.7}\%    & 93.3\%               & 94.1\%                \\
            \midrule
            \multirow{3}{*}{Indoor}  & RandF    & \bf{89.8}\%    & 82.5\%               & 86.4\%                \\
                                     & LightGBM & \bf{92.4}\%    & 87.0\%               & 89.3\%                \\
                                     & SVM      & \bf{94.5}\%    & 91.0\%               & 87.6\%                \\
            \bottomrule
        \end{tabular}
    }\\
    \subfloat[][Position accuracy: $(r,\theta)$.]{
        \begin{tabular}{ccccc}
            \toprule
                                     & ML model & Bi-directional & $\bm{V}^\mathrm{AP}$ & $\bm{V}^\mathrm{STA}$ \\
            \midrule
            \multirow{3}{*}{Outdoor} & RandF    & \bf{74.4}\%    & 65.2\%               & 59.0\%                \\
                                     & LightGBM & \bf{86.1}\%    & 81.0\%               & 77.2\%                \\
                                     & SVM      & \bf{95.2}\%    & 90.5\%               & 90.0\%                \\
            \midrule
            \multirow{3}{*}{Indoor}  & RandF    & \bf{85.9}\%    & 76.0\%               & 79.3\%                \\
                                     & LightGBM & \bf{88.9}\%    & 82.8\%               & 84.0\%                \\
                                     & SVM      & \bf{94.5}\%    & 88.1\%               & 81.4\%                \\
            \bottomrule
        \end{tabular}
    }
\end{table}

\noindent
\textbf{Effect of equipment.}
Table~\ref{tab:humanlocalization_ASUS} summarizes the effect of the equipment on position accuracy in the outdoor environment.
Regardless of the ML model and equipment,
$\bm{V}^\mathrm{AP}$ sensing achieved higher accuracy than $\bm{V}^\mathrm{STA}$ sensing, and the accuracy of bi-directional sensing was higher than that of uni-directional sensing.
Thus, we can conclude that regardless of the equipment, BFM disparity exists in terms of sensing accuracy, and bi-directional sensing is superior to uni-directional sensing.
This further validates the contributions of ~\ref{cnt:bdb_sens_dif_human} and \ref{cnt:bdb_acc} in section~\ref{sec:introduction}.

\begin{table}[t]
    \centering
    \caption{
        Effect of experimental equipment on position accuracy of human localization in outdoor environment.
    }
    \label{tab:humanlocalization_ASUS}
    \begin{tabular}{cccccc}
        \toprule
        Equipment set      & ML model & Bi-directional & $\bm{V}^\mathrm{AP}$ & $\bm{V}^\mathrm{STA}$ \\
        \midrule
        \multirow{3}{*}{A} & RandF    & \bf{74.4}\%    & 65.2\%               & 59.0\%                \\
                           & LightGBM & \bf{86.1}\%    & 81.0\%               & 77.2\%                \\
                           & SVM      & \bf{95.2}\%    & 90.5\%               & 90.0\%                \\
        \midrule
        \multirow{3}{*}{B} & RandF    & \bf{85.2}\%    & 80.8\%               & 62.7\%                \\
                           & LightGBM & \bf{90.2}\%    & 86.7\%               & 78.3\%                \\
                           & SVM      & \bf{92.0}\%    & 88.9\%               & 82.5\%                \\
        \bottomrule
    \end{tabular}
\end{table}

\begin{figure*}[t!]
    \centering
    \subfloat[RandF]{%
        \includegraphics[width=0.3\textwidth]{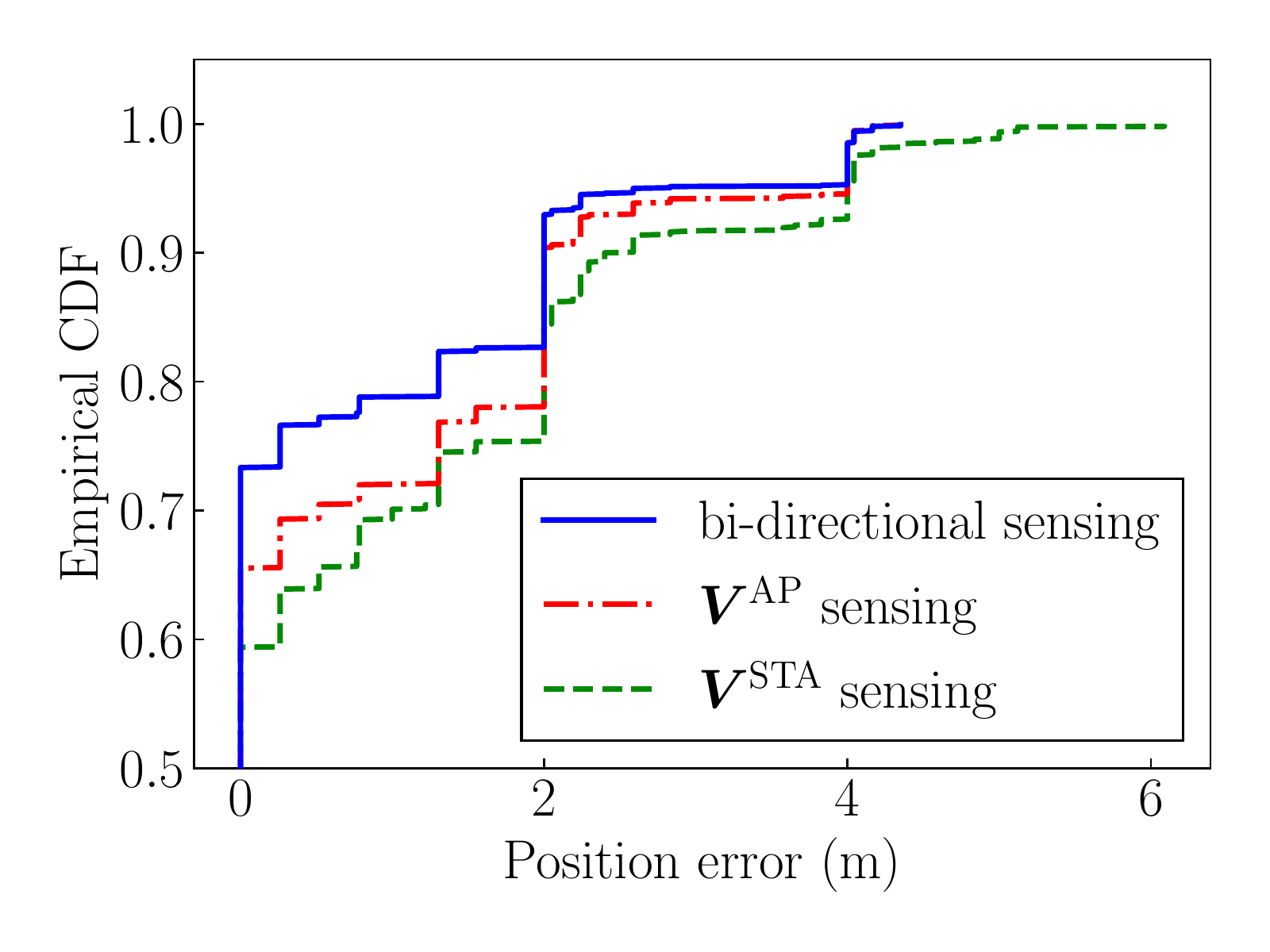}
    }
    \subfloat[LightGBM]{%
        \includegraphics[width=0.3\textwidth]{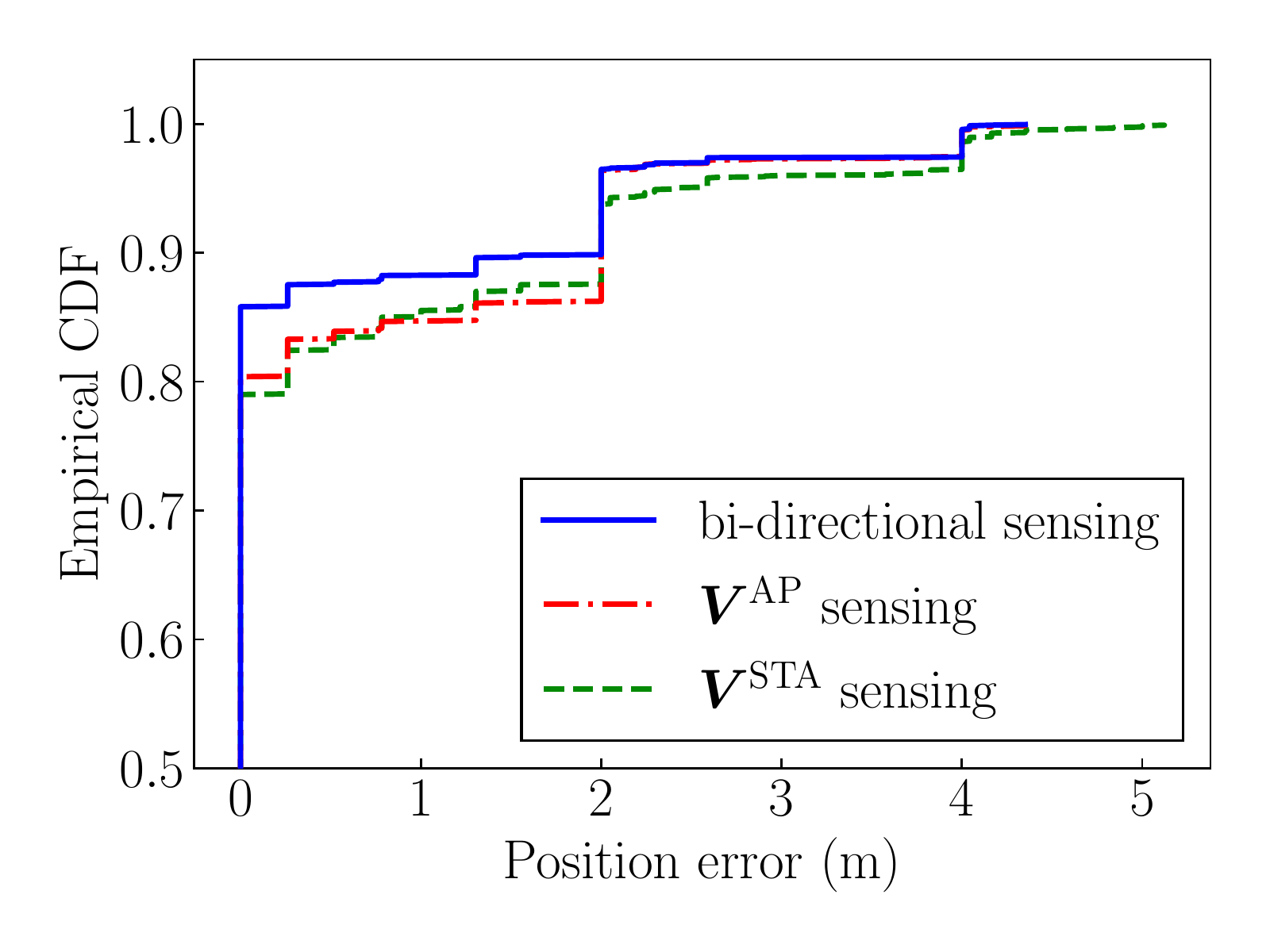}
    }
    \subfloat[SVM]{%
        \includegraphics[width=0.3\textwidth]{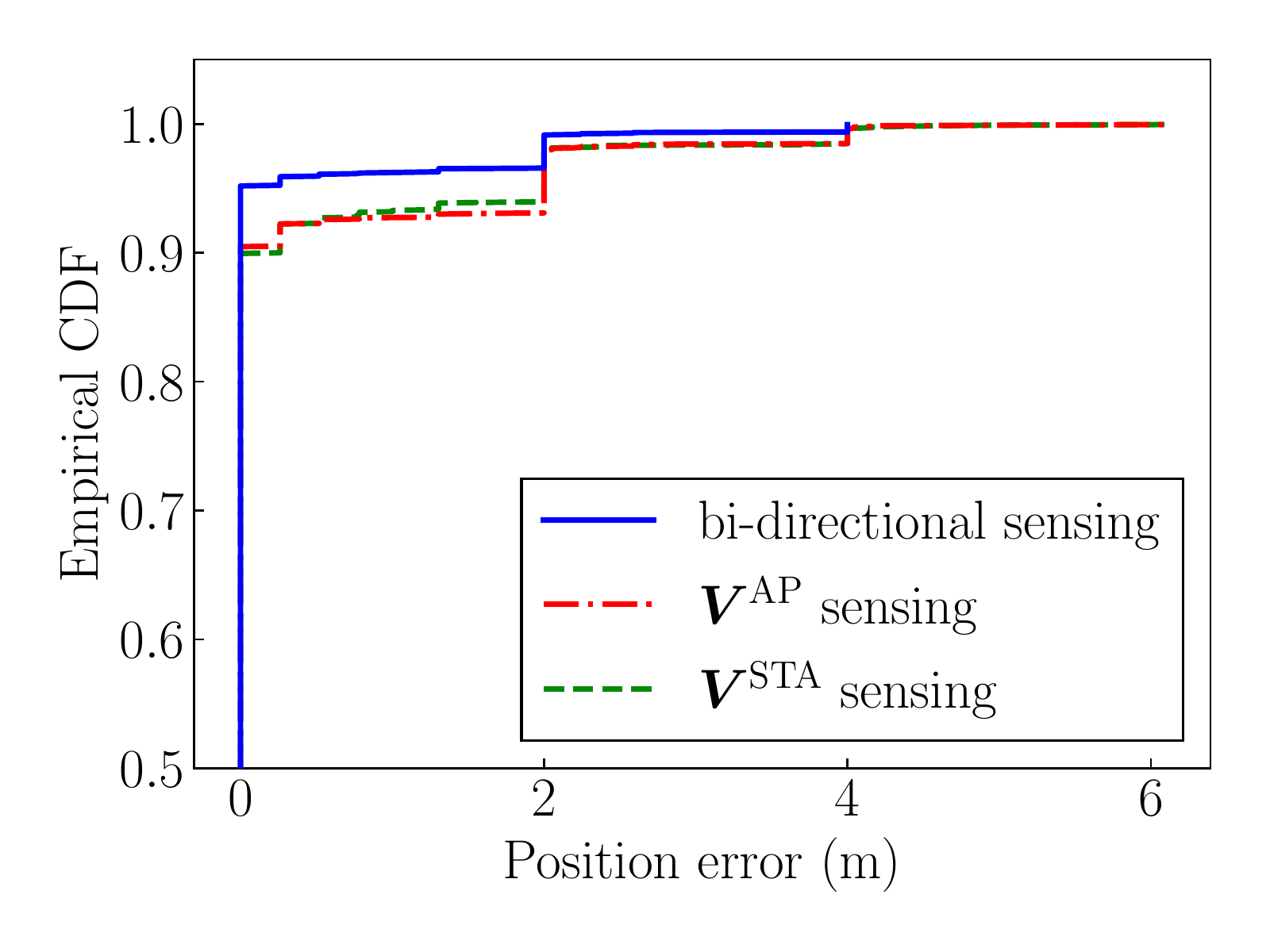}
    }
    \caption{
        Empirical CDF of the localization error for three sensing methods using three ML models.
    }
    \label{fig:human_loc_error_hist}
\end{figure*}

\noindent\textbf{Localization error comparison.}
This section validates that the proposed bi-direction sensing achieves lower human-localization error than uni-directional sensing.
Table~\ref{tab:human_loc_error} shows the average error of human localization tasks in the outdoor environment, wherein the error is defined as the Euclidean distance between the estimated and ground-truth locations.
Regardless of the ML model, bi-directional sensing achieved a lower average error compared to uni-directional sensing.
For example, when using the SVM model,
the average error of bi-directional sensing is lower than 0.1\,m, whereas that of uni-directional sensing is larger than 0.15\,m.
Fig.\ref{fig:human_loc_error_hist} shows the empirical CDF of the human-localization error of three ML models.
Comparing the ratio of the test sample, which has an error less than 1\,m, that of bi-directional sensing is higher compared to that of uni-directional sensing.
For example, when the RandF model is used, the errors associated with bi-directional sensing, $\bm{V}^\mathrm{AP}$ sensing, and $\bm{V}^\mathrm{STA}$ sensing are
74.4\%, 65.2\%, and 59.0\%, respectively.
Thus, we can conclude that bi-direction sensing achieves higher accuracy than uni-directional sensing,
which is consistent with the results presented in this section and further validates \ref{cnt:bdb_acc} in Section~\ref{sec:introduction}.

\begin{table}[t]
    \centering
    \caption{
        Average localization error for the human localization task in outdoor environment.
    }
    \label{tab:human_loc_error}
    \begin{tabular}{cccc}
        \toprule
                 & \hspace{-0.5cm} Bi-directional & $\bm{V}^\mathrm{AP}$ & $\bm{V}^\mathrm{STA}$ \\
        \midrule
        RandF    & \bf{0.526}\,m                  & 0.677\,m             & 0.824\,m              \\
        LightGBM & \bf{0.289}\,m                  & 0.369\,m             & 0.399\,m              \\
        SVM      & \bf{0.090}\,m                  & 0.184\,m             & 0.177\,m              \\
        \bottomrule
    \end{tabular}
\end{table}

\noindent\textbf{Feature importance comparison.}
Table~\ref{tab:feature_importance} shows the feature importance of the RandF and LightGBM models that were trained using bi-directional BFMs.
The feature importance is defined in decision tree models such as the RandF and LightGBM models.
This parameter is assigned to each feature element, and indicates the contribution of each feature to the reduction of the Gini coefficient. A higher importance indicates a greater contribution of the corresponding feature.
Since bi-directional sensing uses the input feature of $\bm{V}^\mathrm{AP}$ and $\bm{V}^\mathrm{STA}$,
Table~\ref{tab:feature_importance} represents the importance assigned to the feature generated from $\bm{V}^\mathrm{AP}$ and that from $\bm{V}^\mathrm{STA}$.
Note that since the target vector is two-dimensional (i.e., angle and distance),
the ML model includes two groups of trees (i.e., angle estimation trees and distance estimation trees);
Thus, we show the feature importance for the two tree groups.

Considering the angle estimation trees, the input features of $\bm{V}^\mathrm{AP}$ have greater importance than those of $\bm{V}^\mathrm{STA}$.
Recall that in terms of the sensing accuracy difference discussed so far, $\bm{V}^\mathrm{AP}$ sensing achieves a higher angle estimation accuracy compared to $\bm{V}^\mathrm{STA}$ sensing.
This importance difference implies that the ML model recognizes that $\bm{V}^\mathrm{AP}$ is more valuable than the $\bm{V}^\mathrm{STA}$.
However, considering the angle estimation trees, the input features of $\bm{V}^\mathrm{AP}$ have comparable or less importance than those of $\bm{V}^\mathrm{STA}$.
This is because the accuracy of $\bm{V}^\mathrm{STA}$ sensing is comparable to or higher than that of $\bm{V}^\mathrm{AP}$ sensing.

\begin{table}[t!]
    \centering
    \caption{
        Feature importance of the ML model trained using bi-directional BFMs.
        The importance of the feature generated from the $\bm{V}^\mathrm{AP}$ and that from $\bm{V}^\mathrm{STA}$ are represented separately.
    }
    \label{tab:feature_importance}
    \subfloat[][Angle estimation trees.]{
        \begin{tabular}{cccc}
            \toprule
                     & $\bm{V}^\mathrm{AP}$ & $\bm{V}^\mathrm{STA}$ \\
            \midrule
            RandF    & \bf{0.692}           & 0.308                 \\
            LightGBM & \bf{0.601}           & 0.399                 \\
            \bottomrule
        \end{tabular}
    }
    \\
    \subfloat[][Distance estimation trees.]{
        \centering
        \begin{tabular}{cccc}
            \toprule
                     & $\bm{V}^\mathrm{AP}$ & $\bm{V}^\mathrm{STA}$ \\
            \midrule
            RandF    & \bf{0.548}           & 0.452                 \\
            LightGBM & 0.426                & \bf{0.574}            \\
            \bottomrule
        \end{tabular}
    }
\end{table}

\section{Conclusion}
In this investigation, it was experimentally validated that the sensing accuracy of two cases of sensing using the BFM transmitted for the AP and sensing based on the BFM transmitted for the STA are different for human localization and the AP's AoD estimation tasks.
The results imply that there exist a potential accuracy degradation in uni-directional BFM-based sensing, which uses either BFM transmitted for the AP or BFM transmitted for STA.
To overcome the potential accuracy degradation, we propose a bi-directional BFM sensing, which simultaneously uses BFMs transmitted for the AP and STA.
We experimentally established that the proposed bi-directional BFM sensing achieved higher sensing accuracy than uni-directional BFM sensing.



\begin{IEEEbiography}
    [{\includegraphics[width=1in, height=1.25in, clip, keepaspectratio]{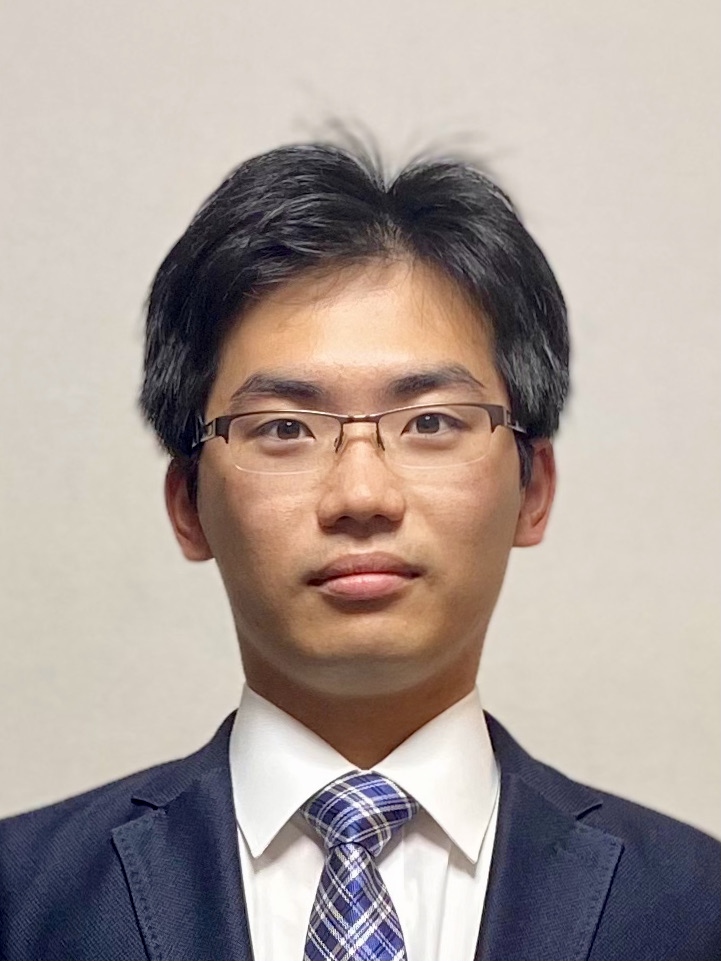}}]
    {Sota~Kondo}
    received the B.E. degree in electrical and electronic engineering from Kyoto University in 2021.
    He is currently studying toward the M.I. degree at the Graduate School of Informatics, Kyoto University.
    He is a student member of the IEEE.
\end{IEEEbiography}

\begin{IEEEbiography}
    [{\includegraphics[width=1in, height=1.25in, clip, keepaspectratio]{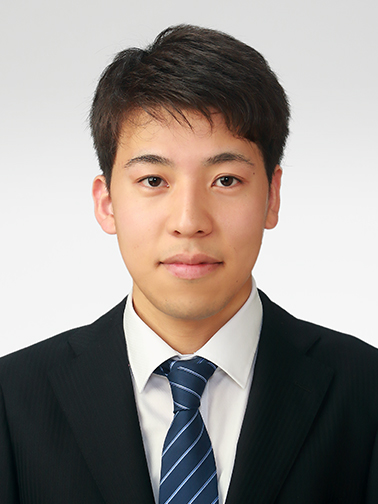}}]
    {Sohei~Itahara}
    received the B.E. degree in electrical and electronic engineering from Kyoto University in 2020.
    He is currently studying toward the M.I. degree at the Graduate School of Informatics, Kyoto University.
    He is a student member of the IEEE.
\end{IEEEbiography}

\begin{IEEEbiography}
    [{\includegraphics[width=1in, height=1.25in, clip, keepaspectratio]{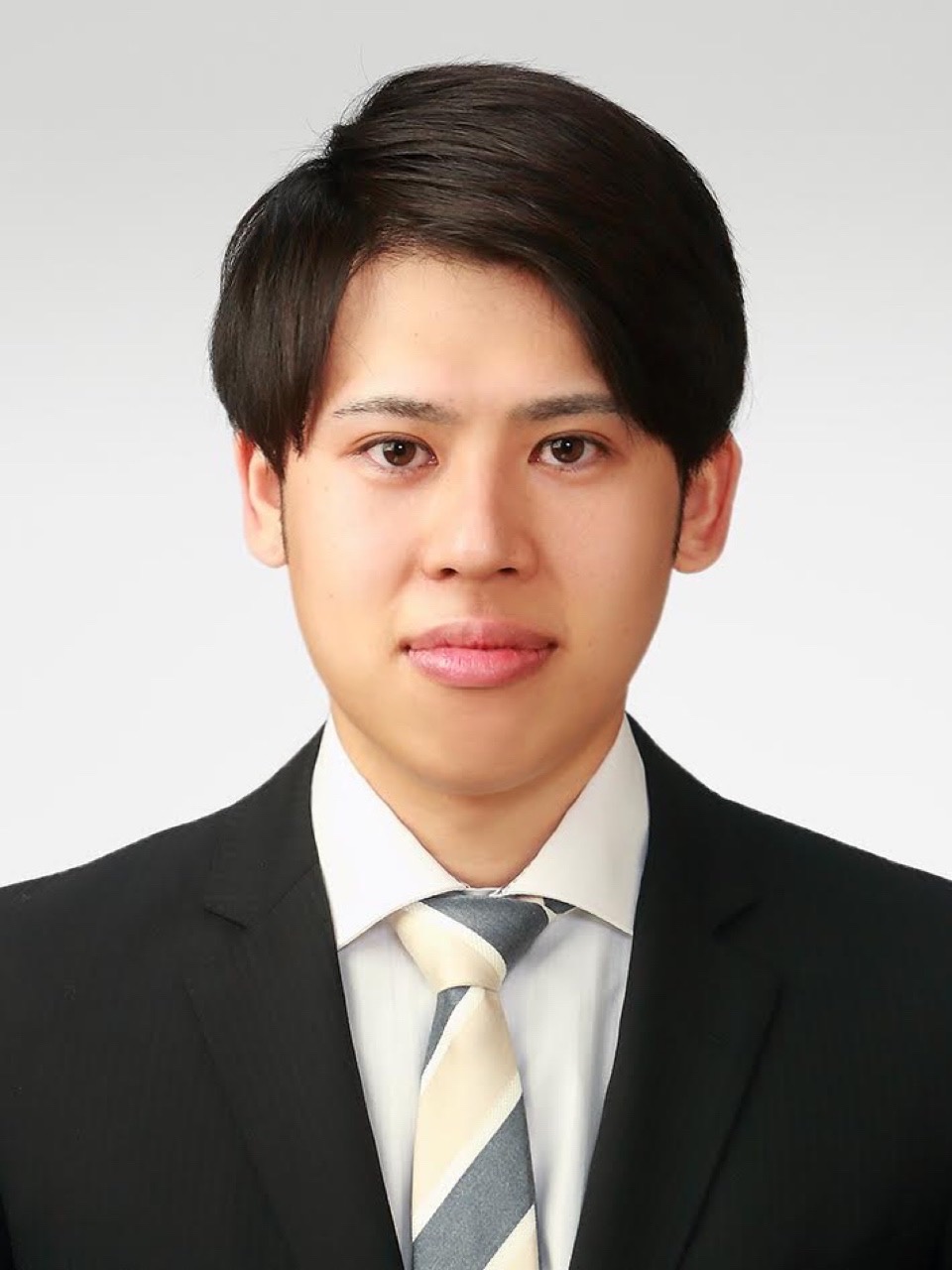}}]
    {Kota~Yamashita}
    received the B.E. degree in electrical and electronic engineering from Kyoto University in 2020.
    He is currently studying toward the M.I. degree at the Graduate School of Informatics, Kyoto University.
    He is a student member of the IEEE.
\end{IEEEbiography}

\begin{IEEEbiography}
    [{\includegraphics[width=1in, height=1.25in, clip, keepaspectratio]{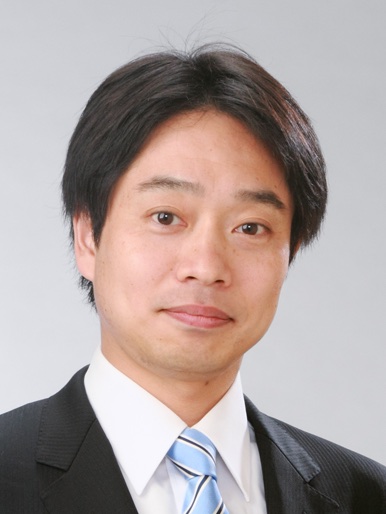}}]
    {Koji~Yamamoto}
    received the B.E.\ degree in electrical and electronic engineering from Kyoto University in 2002, and the master and Ph.D.\ degrees in Informatics from Kyoto University in 2004 and 2005, respectively.
    From 2004 to 2005, he was a research fellow of the Japan Society for the Promotion of Science (JSPS).
    Since 2005, he has been with the Graduate School of Informatics, Kyoto University, where he is currently an associate professor.
    From 2008 to 2009, he was a visiting researcher at Wireless@KTH, Royal Institute of Technology (KTH) in Sweden.
    He serves as an editor of IEEE Wireless Communications Letters, IEEE Open Journal of Vehicular Technology, and Journal of Communications and Information Networks, a symposium co-chair of GLOBECOM 2021, and a vice co-chair of IEEE ComSoc APB CCC.
    He was a tutorial lecturer in IEEE ICC 2019.
    His research interests include radio resource management, game theory, and machine learning.
    He received the PIMRC 2004 Best Student Paper Award in 2004, the Ericsson Young Scientist Award in 2006.
    He also received the Young Researcher's Award, the Paper Award, SUEMATSU-Yasuharu Award, Educational Service Award from the IEICE of Japan in 2008, 2011, 2016, and 2020, respectively, and IEEE Kansai Section GOLD Award in 2012.
    He is a senior member of the IEEE and a member of the Operations Research Society of Japan.
\end{IEEEbiography}

\begin{IEEEbiography}[{\includegraphics[width=1in, height=1.25in, clip, keepaspectratio]{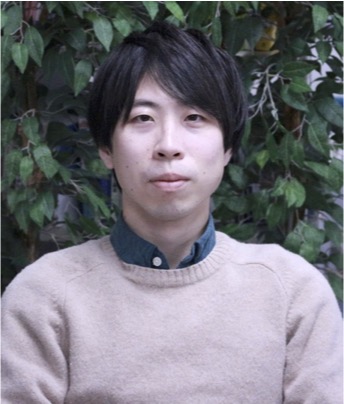}}]{Yusuke~Koda}
    received the B.E. degree in electrical and electronic engineering from Kyoto University in 2016, and the M.E. and the Ph.D. degree in informatics from the Graduate School of Informatics, Kyoto University in 2018 and 2021, respectively. He is currently a Postdoctoral Researcher with the Centre for Wireless Communications, University of Oulu, Finland, where he visited the Centre for Wireless Communications in 2019, to conduct collaborative research. He received the VTS Japan Young  Researcher's Encouragement Award in 2017 and TELECOM System Technology Award in 2020. He was a recipient of the Nokia Foundation Centennial Scholarship in 2019.
\end{IEEEbiography}

\begin{IEEEbiography}
    [{\includegraphics[width=1in, height=1.25in, clip, keepaspectratio]{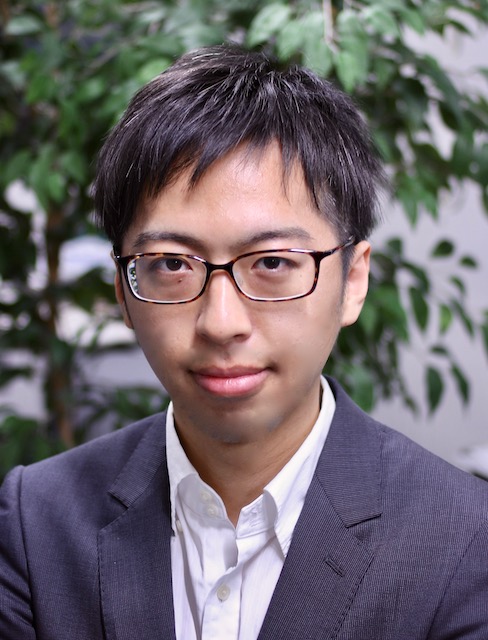}}]
    {Takayuki~Nishio} received the B.E.\ degree in electrical and electronic engineering and the master's and Ph.D.\ degrees in informatics from Kyoto University in 2010, 2012, and 2013, respectively.
    He was an assistant professor in communications and computer engineering with the Graduate School of Informatics, Kyoto University from 2013 to 2020. From 2016 to 2017, he was a Visiting Researcher with the Wireless Information Network Laboratory (WINLAB), Rutgers University, USA. Since 2020, he has been an Associate Professor with the School of Engineering, Tokyo Institute of Technology, Japan.
    Wireless Information Network Laboratory (WINLAB), Rutgers University, United States. 
    His current research interests include machine learning-based network control, machine learning in wireless networks, vision-aided wireless communications, and heterogeneous resource management.
\end{IEEEbiography}

\begin{IEEEbiography}
    [{\includegraphics[width=1in, height=1.25in, clip, keepaspectratio]{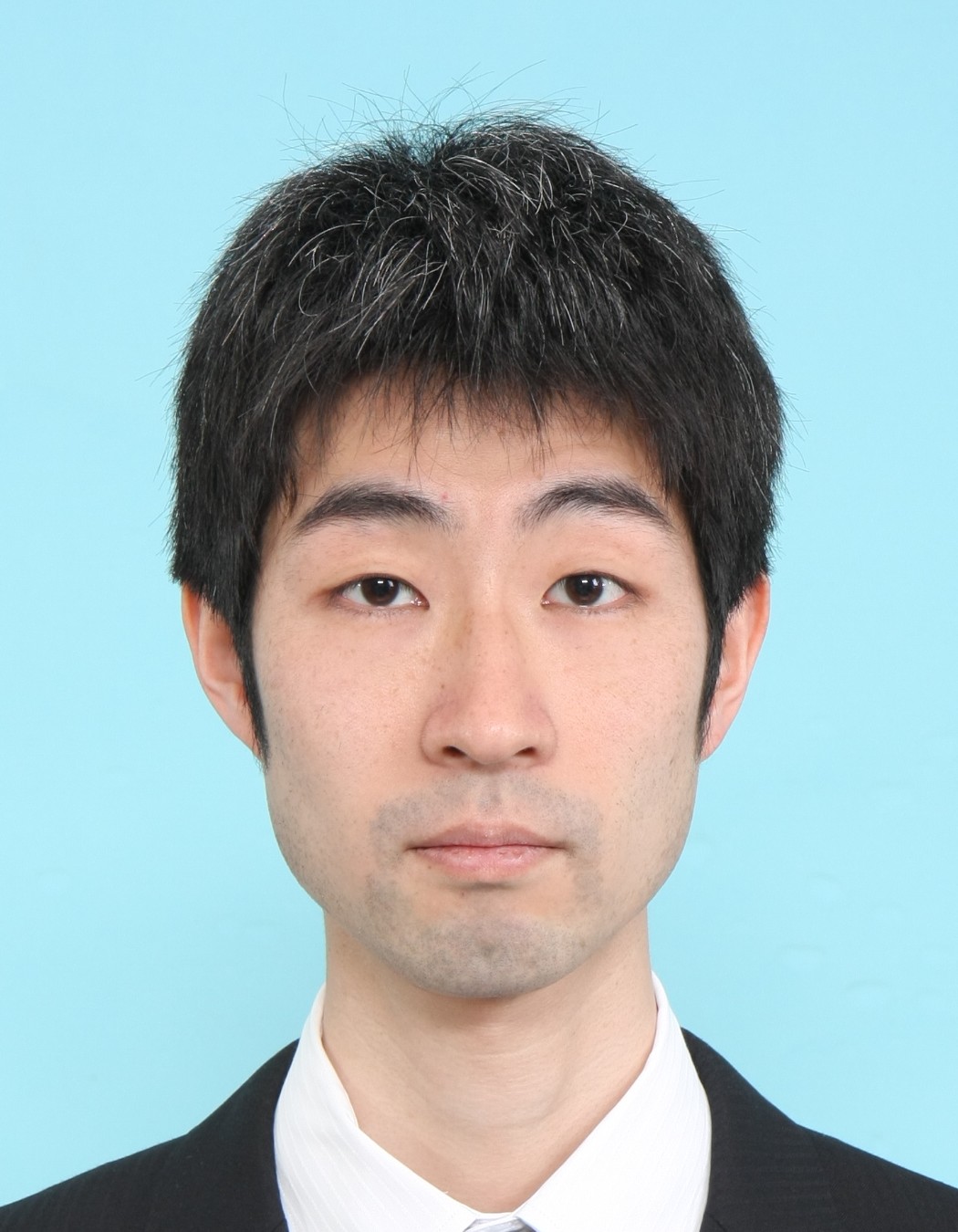}}]
    {Akihito~Taya} received the B.E. degree in electrical and electronic engineering from Kyoto University, Kyoto, Japan in 2011,
    and the master and Ph.D.\ degree in Informatics from Kyoto University in 2013 and 2019, respectively.
    From 2013 to 2017, he joined Hitachi, Ltd., where he participated in the development of computer clusters.
    He has been an assistant professor of the Aoyama Gakuin University, since 2019.
    He received the IEEE VTS Japan Young Researcher's Encouragement Award and the IEICE Young Researcher's Award in 2012 and 2018, respectively.
    His current research interests include distributed machine learning and human activity and emotion recognition using sensor networks.
    He is a member of the IEEE.
\end{IEEEbiography}

\EOD
\end{document}